\documentclass{jfm}
\ifCUPmtlplainloaded \else %
  \checkfont{eurm10} %
  \iffontfound %
    \IfFileExists{upmath.sty} %
      {\typeout{^^JFound AMS Euler Roman fonts on the system, %
                   using the 'upmath' package.^^J}
       \usepackage{upmath}} %
      {\typeout{^^JFound AMS Euler Roman fonts on the system, but you %
                   dont seem to have the}
       \typeout{'upmath' package installed. JFM.cls can take advantage %
                 of these fonts,^^Jif you use 'upmath' package.^^J}
      } %
  \else %
  \fi %
\fi %
\ifCUPmtlplainloaded \else %
  \checkfont{msam10} %
  \iffontfound %
    \IfFileExists{amssymb.sty} %
      {\typeout{^^JFound AMS Symbol fonts on the system, using the %
                'amssymb' package.^^J}
       \usepackage{amssymb}
       \let\le=\leqslant \let\leq=\leqslant %
       \let\ge=\geqslant \let\geq=\geqslant %
      }{} %
  \fi %
\fi %
\ifCUPmtlplainloaded \else %
  \IfFileExists{amsbsy.sty} %
    {\typeout{^^JFound the 'amsbsy' package on the system, using it.^^J}%
     \usepackage{amsbsy}} %
    {\providecommand\boldsymbol[1]{\mbox{\boldmath $##1$}}} %
\fi %

\usepackage{color}
\usepackage{amsmath}
\usepackage{amssymb}
\usepackage{epsfig}
\usepackage{graphicx}
\usepackage{psfrag}

\usepackage{natbib}
\usepackage{pifont}

\newcommand\etal{\mbox{\textit{et al.\ }}}

\newcommand{\dd}[2]{\frac{d^2 #1}{d #2^2}}

\renewcommand{\u}{{\mathbf{u}}}
\newcommand{\uz}{{\mathbf{u}}_{000}}
\newcommand{\uk}{{\mathbf{u}}_{kn\omega}}
\newcommand{\f}{{\mathbf{f}}}
\newcommand{\fz}{\f_{000}}
\newcommand{\fk}{\f_{kn\omega}}
\newcommand{\Lk}{{\mathcal{L}}_{kn\omega}}
\renewcommand{\k}{kn\omega}

\def\BibTeX{{\rm B\kern-.05em{\sc i\kern-.025em b}\kern-.08em
    T\kern-.1667em\lower.7ex\hbox{E}\kern-.125emX}}

\def\grad{\V{\nabla}}

\def\div{\grad \cdot }

\def\s={\stackrel{s}{=}}

\newcommand{\ud}{{\mathrm{d}}}

\newcommand{\curly}[1]{{\cal{#1}}}

\newcommand{\V}[1]{{\boldsymbol{#1}}}

\newcommand{\bspace}[1]{{\mathbb{#1}}}

\newcommand{\norm}[2]{\lVert#1\rVert_{#2}}
\newcommand{\inprod}[2]{\left({#1},{#2}\right)}

\title[A critical layer model for turbulent pipe flow]{A critical layer model for turbulent pipe flow}
\author[B. J. McKeon and A. S. Sharma]
{B.\ns J.\ns M\ls c\ls K\ls E\ls O\ls N$^1$\ls\ns and A.\ns S.\ns S\ls H\ls A\ls R\ls M\ls A$^2$\ls}

\affiliation{$^1$Graduate Aerospace Laboratories, \\California Institute of Technology, Pasadena, CA 91125, USA\\[\affilskip]
$^2$Department of Aeronautics, \\Imperial College, London SW7 2AZ, UK}

\date{\today}

\pubyear{2009} 
\volume{???} 
\begin{document}
\maketitle

\begin{abstract}

A model-based description of the scaling and radial location of turbulent fluctuations in turbulent pipe flow is presented and used to illuminate the scaling behaviour of the very large scale motions. The model is derived by treating the nonlinearity in the perturbation equation (involving the Reynolds stress) as an unknown forcing, yielding a linear relationship between the velocity field response and this nonlinearity. We do not assume small perturbations. We examine propagating modes, permitting comparison of our results to experimental data, and identify the steady component of the velocity field that varies only in the wall-normal direction as the turbulent mean profile. The ``optimal'' forcing shape, that gives the largest velocity response, is assumed to lead to modes that will be dominant and hence observed in turbulent pipe flow.

An investigation of the most amplified velocity response at a given wavenumber-frequency combination reveals critical layer-like behaviour reminiscent of the neutrally stable solutions of the Orr-Sommerfeld equation in linearly unstable flow. Two distinct regions in the flow where the influence of viscosity becomes important can be identified, namely a wall layer that scales with $R^{+1/2}$ and a critical layer, where the propagation velocity is equal to the local mean velocity, that scales with $R^{+2/3}$ in pipe flow. This framework appears to be consistent with several scaling results in wall turbulence and reveals a mechanism by which the effects of viscosity can extend well beyond the immediate vicinity of the wall.

The model reproduces inner scaling of the small scales near the wall and an approach to outer scaling in the flow interior. The appropriate scaling velocity for the very large scale motions is predicted to be the centreline velocity, in agreement with experimental results. Lastly, we interpret the wall modes as the motion required to meet the wall boundary condition, identifying the interaction between the critical and wall modes as a potential origin for an interaction between the large and small scales that has been observed in recent literature as an amplitude modulation of the near-wall turbulence by the very large scales.

\end{abstract} 

\section{Introduction}
The accurate description of statistical scaling and instantaneous structural coherence of turbulent fluctuations, and their relationship to the mean flow, are amongst the important unsolved problems in physics.
Even at the simplest level, we still lack a complete explanation of the development of the mean velocity profile in canonical flows as the Reynolds number increases. Such an understanding would be an important step towards accurate prediction of skin friction in complex aeronautical and industrial flows, and would underpin any effort at a turbulence control scheme. A brief summary of three distinct approaches to these problems is given here.

The research literature is replete with statistical descriptions of the wall-normal distributions of the components of the Reynolds stress tensor, with an obvious bias to the streamwise normal stress, which is most easily measured.  New understanding of turbulence has emerged in the past decade. This has included differences in some characteristics between the canonical cases of pipe, channel and boundary layer flow~\citep{Monty09}, in contrast to the long-standing hypothesis of universality of near-wall scaling. Relevant to the present work, several studies have revealed highly energetic structures with streamwise wavelength of order ten times the outer lengthscale, deemed variously \emph{Very Large Scale Motions} (VLSMs) or \emph{superstructures}.
The VLSM phenomenon, discussed at length later in this work, suggests a more complicated nature of the scaling of the turbulence than an inner/outer/overlap structure proposed for the mean velocity. This has been confirmed to the resolution of experimental measurements, for the mean velocity. Specifically, it confirms the influence on the inner, near-wall flow, of turbulent activity that scales on outer variables.

The advent of Particle Image Velocimetry (PIV) techniques and advances in simulation, alongside more traditional visualisation techniques, have illuminated the development and grouping of organised, coherent structures. These include the autonomous near-wall cycle and an eddy type that is statistically well described by the hairpin vortex paradigm.

In parallel to these experimental and computational approaches, considerable progress has been made in understanding the amplification properties of the Navier-Stokes operator, and in particular the linearised operator, in laminar flows.  More recent work has extended some of these techniques to the turbulent case, with limited success.

In what follows, we summarise some key concepts and questions arising from these three distinct approaches which are pertinent to the current work.

\subsection{The challenge to classical scaling: the influence of VLSMs on the near-wall region}

Classical scaling ideas involve inner and outer layers, where the appropriate spatial scales are respectively the viscous unit and the outer lengthscale, and the appropriate velocity scale is the friction velocity. In the case of the outer layer this velocity is impressed by the boundary condition at the wall. For sufficiently high Reynolds number, there may also be an overlap layer in which both scalings hold and therefore the important lengthscale must be the local distance from the wall. This scaling appears to work well for the mean velocity despite recent challenges \citep{Marusic09}. However it is unable to collapse the turbulent fluctuations in the inner and overlap regions, an apparent reflection of the influence of outer scales on the inner region, termed ``inner-outer interaction'' as explored by~\cite{Bandy84}, that is reflected in the spectral energy distribution.

The recent focus on VLSMs in canonical turbulent wall flows has given some insight into the source of this interaction. Coherence across the wall layer at long streamwise lengthscales has been known since the observations of \cite{Kovasznay70}. The work of \cite{Kim00} and \cite{Morrison04} identified the energetic importance of very large scale motions in the streamwise spectra in turbulent pipe flow and the subsequent work by Adrian and co-workers~\citep{Guala06,BalakumarPTRSA} has indicated that large and very large scale features must be considered to be ``active'' in the sense that they carry significant shear stress. This contrasts with Townsend's geometrical arguments that only the small scales display sufficient coherence in the wall-normal and streamwise fluctuations to extract significant energy from the mean flow, via the product of the Reynolds shear stress and the mean velocity gradient.

Simulating and observing these high aspect ratio VLSMs places heavy demands on existing computational and experimental techniques, a constraint that has obstructed progress in our understanding of their origin and development in turbulent flows.
The statistical imprint in the streamwise and wall-normal directions is clear from hot-wire measurements. \cite{HutchinsPTRSA07} showed for sufficiently high Reynolds numbers that this imprint reaches from a peak energy located somewhere in the overlap region down to the wall, even having a footprint on the wall shear stress~\citep{Marusictau07}: \cite{Hutchins07} and \cite{Monty07} used Taylor's hypothesis to reconstruct the long streamwise-spanwise coherence in the streamwise velocity from arrays of hot-wires in turbulent boundary layers and pipe flow, respectively. Their use of Taylor's hypothesis highlights a major observational difficulty: at these large scales, Taylor's hypothesis may not apply, since the necessary arguments based on the ratio of eddy turnover timescale to convective timescale no longer hold.
In addition, the wall-normal extent of the VLSMs means that if the VLSM convects with the local mean velocity somewhere in the overlap layer, Taylor's hypothesis with a scale-independent convective velocity must be in error closer to the wall, and likely also further from the wall to a much lesser degree.

Having listed some of the outstanding questions concerning the VLSMs, we now provide a brief summary of what is known about their extent and influence on the near-wall region. Following the first observation by Hutchins \& Marusic that the peak streamwise VLSM energy appeared to occur at a constant fraction of the boundary layer thickness, $y/\delta \sim 0.05$, the location of the peak has been shown by \cite{mckeonAIAA08} and \cite{Mathis09} to have a weak Reynolds-number dependence when a sufficiently large range of Reynolds number considered. The latter authors compared the peak's location with the centre of a mean velocity overlap region that has an inner limit that is either fixed in inner units or Reynolds number dependent, giving rise to $\delta^{+ 1/2}$ and $\delta^{+ 3/4}$ dependence of the peak location, respectively, where $\delta^+ = \delta u_\tau/\nu$, $\nu$ is the kinematic viscosity and the friction velocity $u_\tau = \sqrt{\tau/\rho}$. However the agreement depends crucially on single data points obtained at very high Reynolds number in the near-neutral atmospheric surface layer, and these vary between studies~\citep{Mathis09,Metzger07,Guala09mod}.

Recent studies show that the streamwise extent of the dominant large scale motion is proportional to the outer flow lengthscale and is closely given by $\lambda_{x,VLSM} \sim 10R$ in internal flows~\citep{Kim99,Monty09} and $\lambda_{x,superstructure} \sim 6\delta$ in boundary layers~\citep{HutchinsPTRSA07}, while the Direct Numerical Simulation (DNS) of channel flow  of \cite{Jimenezlargescale04} and the study of Monty \etal indicate the additional importance of a slightly smaller wavelength, $\lambda_{x,LSM} \sim 3h$. Here $R,\delta$ and $h$ are the pipe radius, boundary layer thickness and channel half-height, respectively. \cite{Monty09} recently performed a rigorous comparison of the streamwise velocity spectra in the different flows at the same Reynolds number and elucidated the respective importance of these three wavelengths in different regions of the flow. The spanwise extent in each case appears to be of the order of one outer lengthscale~\citep{Hutchins07,Monty07}, giving the largest scale structure an approximate axially elongated aspect ratio of $10:1:1$.

Subsequently \cite{Mathis09} expounded on earlier observations that the very large scales apply an amplitude modulation to the small scale turbulence near the wall.  Using a Hilbert transform technique, they were able to quantify the interaction and its Reynolds number dependence, as well as demonstrate a change in the sign of the modulation that corresponded with the location of the VLSM energy peak. \cite{Guala09sf}, \cite{Guala09mod} and \cite{Chung09} have also proposed methodologies to describe this effect, with the latter identifying that the modulation can also be described in terms of the spatial phase relationship between large and small scale turbulent activity.

It is clear, then, that the very large scale structure reflects an aspect of boundary layer dynamics that has hitherto been poorly understood, but has importance for scaling of local and global turbulence properties, as well as for future flow control schemes.

\subsection{Alternative scaling approaches from theory and observation: Critical Layers}

Several theories have been proposed to account for the missing physics in the classical scaling, including the mesolayer of \cite{LongChen81}, the focus on the location of the Reynolds stress peak by Sreenivasan and co-workers \citep{Sreeni97} and the hierarchical structure associated with the mean momentum balance of Klewicki and co-workers~\citep{Klewicki07}.

\cite{Sreenivasan88} formulated an inviscid structural model of the turbulent boundary layer consisting of two vortex sheets of opposite sign symmetrically located about the hypothetical wall location.  By analogy with the critical layer in transitional boundary layers, he showed that the wall-normal location of the peak in Reynolds shear stress should have a Reynolds number dependence given by $y^+_{pk} \sim R^{+ 1/2}$ with a proportionality constant of two determined from experimental data. Here $y^+ = y u_\tau/\nu$ is the wall-normal distance $y$ non-dimensionalised with the viscous scaling length, defined as the kinematic viscosity $\nu$ divided by the friction velocity. The wall shear stress is denoted $\tau_w$ and $\rho$ represents the density.
Experimental evidence from boundary layers, pipe and channel flows supported this critical layer interpretation, the resulting streamwise and spanwise wavelengths for maximum amplification,  the $y^+_{pk} \sim R^{+ 1/2}$ scaling~\citep{Sreeni97} and the hypothesis that the mean velocity at the peak in the Reynolds shear stress corresponds to a constant fraction of the freestream velocity.  In addition, \cite{Sreeni06} were able to extend the study of the region in the vicinity of the Reynolds stress peak by performing a logarithmic expansion and predicting the resultant form for the mean velocity profile.

The reader should note that an analogy with critical layer theory in laminar flow was also made by \cite{Sirovich90}, who found that the most energetic propagating eigenfunctions in a Proper Orthogonal Decomposition\footnote{also known as Karhunen-Loeve analysis} (POD) of turbulent channel flow had their principal support in the region of the peak in Reynolds shear stress. \cite{Duggleby07} have also given some  insight into the most energetic POD modes in low Reynolds number turbulent pipe flow, illuminating a distinction between propagating and non-propagating, wall, lift, asymmetry and ring modes defined by the relative magnitude of the streamwise and azimuthal wavenumbers.  A truncated POD representation of turbulent pipe flow was also used by \cite{Aubry88} to develop a model for the dynamical behaviour of the streamwise roll modes and establish a connection between near-wall turbulent flow and the dynamics of a chaotic system.

\subsection{Linear and nonlinear tools for analysis of the Navier-Stokes equations}

Other researchers, essentially forming a different community, have exploited tools from linear systems and dynamical systems theory, with a notable focus on the large algebraic energy growth that is possible due to the non-normality of a stable Navier-Stokes operator that has been linearised about a base flow. In this picture of transition, for sufficiently large perturbations, such growth is understood to then bring the locally stable system out of its basin of attraction, inducing nonlinear behaviour that is associated with turbulence.
An operator $A$ is non-normal if $AA^*\neq A^*A$ where $A^*$ denotes the adjoint of $A$.  The adjoint is defined with respect to an inner product, so we see that non-normality is only defined with respect to a particular inner product. Typically, for this type of study, the $\mathcal{L}_2$ or perturbation energy norm is of interest. Flows that are linearised about a steady flow solution with shearing can yield highly non-normal operators. In the past fifteen years, significant progress has been made in understanding system non-normality as a mechanism for energy amplification in shear flows~\citep{Butler92, Trefethen93, Farrell93} leading to nonlinear breakdown in both linearly stable and unstable flows~\citep{Jovanovicsupercrit04}.

There has been less investigation of the non-normal growth mechanisms in turbulent flow, with the notable exceptions of the early attempt by \cite{Butler92} to predict the spacing of near-wall streaks in turbulent flow and the more recent studies of \cite{delAlamo06}, \cite{Cossu09} and \cite{Willis09}. A notable and somewhat limiting issue in the treatment of turbulent flow is the modelling of the interaction of the amplified disturbances with the ``background'' turbulence. A solution introduced by \cite{Reynolds72} has been to use the eddy viscosity formulation of \cite{Cess58}, but this relies on an \emph{a priori} knowledge of the spatially-averaged, wall-normal variation of the mean Reynolds stress integrated across contributions from various Reynolds numbers. Other attempts have been made to use linear analysis to explain the dominant features of turbulent flow in terms of optimal transient modes in the initial value problem~\citep{Butler92, delAlamo06, Cossu09}, response to stochastic forcing~\citep{Farrell98, Bamieh01} and system norm analysis~\citep{Jovanovic05,Meseguer03}. In recent work, \cite{Willis09} have investigated the maximal response to harmonic forcing in pipe flow. Perhaps most importantly, it has been shown that both linear non-normality~\citep{Henningson94} and the terms that are linear in the turbulent fluctuation~\citep{Kim00} are required to sustain turbulence in infinite or periodic wall-bounded flows.

\cite{delAlamo06} made a direct comparison between a transient growth analysis of channel flow and earlier DNS results, and showed that while the analysis could predict two spanwise wavelengths that would experience large transient energy growth that were in good agreement with the computational and experimental observations, the corresponding predicted streamwise wavelengths were significantly too high.  \cite{Cossu09} performed a similar analysis in a turbulent boundary layer, modelling it as a parallel flow.  The realisation that turbulent wall flows can be linearly stable followed the work of \cite{ReynoldsT67}, who demonstrated this for turbulent channel flow.

In terms of the physical mechanisms and structure behind the energetic near-wall cycle, \cite{Schoppa02} used linear perturbation methods to propose a streak transient growth mechanism capable of reproducing with good fidelity structures observed in a low Reynolds number DNS of a channel. \cite{Waleffe97, Waleffe01, Waleffe03} followed a different approach, developing exact solutions of the Navier-Stokes equations that give rise to unstable coherent structures which they propose form the foundations of transitional flow and near-wall turbulence.  Subsequently there has been much interest in the importance and observability of travelling wave solutions of the Navier-Stokes equations, e.g. \cite{Wedin04} and \cite{Viswanath09} in pipe flow (which, of course, is also linearly stable). The ongoing work of \cite{Gayme2D3C} with a forced, two-dimensional, three velocity component model provides another attempt to predict the form of the turbulent mean flow.

\subsection{Selection of pipe flow for further study}

Of the canonical turbulent flows, pipe flow has received considerable attention since Osborne Reynolds' seminal work identifying the role of the Reynolds number in the flow behaviour. Pipe flow is also important to many obvious industrial applications. From an experimental point of view, providing that the development length is sufficiently long for the flow to be considered to be fully-developed, a pipe constitutes a well-defined, simple to generate geometry.  As such, there is a wealth of experimental pipe flow data available for comparison, with the results from the Princeton/ONR Superpipe providing detailed information on the mean velocity~\citep{ZS,mckeonmean04}, streamwise~\citep{Morrison04} and wall-normal~\citep{Zhao07} fluctuations, and azimuthal correlations \citep{Bailey08,BaileyPOD09} across three decades in Reynolds number.  The results of \cite{Monty07} at an intermediate Reynolds number have already been described in the context of the VLSMs.  In addition, a recent DNS study by \cite{Wu08} explored the properties of the spatial velocity field at relatively low turbulent Reynolds number, while the POD analysis of \cite{Duggleby07} categorised energetic propagating and non-propagating modes \emph{a posteriori}, from full field information from a DNS at $R^+=180$. Lastly, the geometry of the pipe has the useful property of imposing a restriction on the azimuthal wavenumber to integer values only, which simplifies our analysis in the subsequent sections.

\subsection{Contribution}
In this work we propose a simple analysis of the Navier-Stokes equations for incompressible, fully-developed turbulent pipe flow. The resulting model describes the spatial distribution of turbulent energy in the three-dimensional propagating velocity modes that are most responsive to harmonic forcing.
We offer the model as a first step towards bridging the gap between statistical and structural interpretations of wall turbulence, since it provides qualitative information on both the temporal and spatial distributions of velocity associated with each mode. We suggest that such a reconciliation of the differing observations of the same system would provide an important advance in the field. In what follows, we use the model to investigate an issue of intense current interest in the boundary layer community; the radial extent, characteristics and scaling of very large scale motions.

\section{A simple model for the spatio-temporal distribution of turbulent energy in pipe flow}\label{sect:approach}

In following analysis, we develop a formulation of the Navier-Stokes equations designed to examine the receptivity of turbulent pipe flow to forcing.  In this way, we develop a framework that permits investigation of the form and likely magnitude of spatially- and temporally-harmonic, propagating finite-amplitude fluctuations about the turbulent mean profile in pipe flow.

\subsection{Pipe flow equations and non-dimensionalisation}

\begin{figure}
\begin{center}
\psfrag{q,w}[][]{$\theta,w$}
\psfrag{r,v'}[][]{$r,v'$}
\psfrag{x,u}[][]{$x,u$}
\psfrag{y,v}[][]{$y,v$}
\includegraphics[width=10cm]{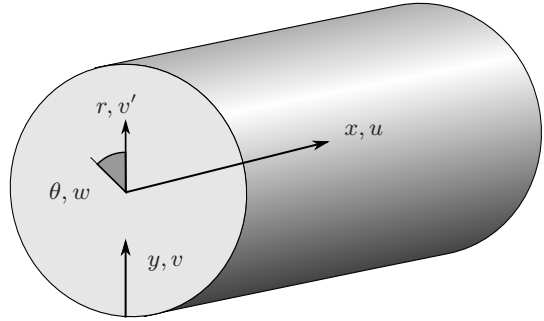}
\caption{Schematic of pipe geometry and nomenclature.}
\label{fig:pipe}
\end{center}
\end{figure}
The non-dimensional Navier-Stokes equations for fully-developed, incompressible pipe flow with constant viscosity are given by
\begin{align}
\partial_t \u =& -\nabla p -\u\cdot\nabla\u  + \frac{1}{Re}\nabla^2 \u\\
\div \u =& 0
\label{eq:pipe}
\end{align}
We follow the convention of \cite{Meseguer03}, where the equations of motion are non-dimensionalised with respect to the pipe diameter and twice the bulk, volume-averaged velocity, $U_{bulk}$ (which in their study is equal to the laminar flow centreline velocity). Thus the Reynolds number in Equation~\ref{eq:pipe} can be defined as
\[Re=\frac{U_{bulk}D}{\nu}.\]
Here $D$ is the pipe diameter and $\nu$ is the kinematic viscosity.  We retain the boundary layer terminology in fixing $y=1-r$, and $u$, $v(=-v')$ and $w$ as corresponding to the streamwise, wall-normal and azimuthal velocities such that $\u=(v',w,u)$, as shown in figure~\ref{fig:pipe}.

\subsection{Model development}

We introduce a projection onto a divergence free basis: $\left\{\xi_m(r)\right\}$ in the radial direction, Fourier modes in the homogeneous spatial directions and the Laplace transform in time.
Implicitly, we are considering a pipe that is infinitely long or periodic in the axial direction.
Assuming fully developed flow allows us to express the velocity field as the sum of harmonic modes.
Then,
\begin{align}
u(r,x,\theta,t) &= \frac{1}{2\pi i} \sum_{m,n}\int_{-i\infty}^{i\infty}\int_{-\infty}^{\infty} c_{mkn\omega} \xi_m(r)e^{ikx+in\theta+st} \ud k \ud s\\
&=\frac{1}{2\pi i} \sum_{m,n}\int_{-i\infty}^{i\infty}\int_{-\infty}^{\infty} c_{mkn\omega} \xi_{mkn\omega}(r)\ud k \ud s
\end{align}
with $s=i\omega$, so that only harmonic forcing and response is considered.

The wavenumbers $(k,n,\omega)$ are non-dimensional such that $k=k'R$, etc.  The integration path for the Laplace transform is over the closed right half plane, which is analytic in the case of pipe flow. In the inviscid limit special treatment for singularities on the imaginary axis would be required, but we do not consider this limit.

We work with the spatial $\curly{L}_2$ inner product throughout,
\begin{equation}
\inprod{a}{b} = \int_{x\in \Omega} a(x) b(x) dx.
\end{equation}

The basis functions are required to have the special properties
\begin{align}
\inprod{\xi_a}{\xi_b}   &= \delta_{ab},\\
\div \xi(r)_{mkn\omega} &= 0.
\end{align}
This is done to eliminate the pressure term. For notational convenience we make the definitions
\begin{align}
\uk        &=\inprod{\u}{e^{i(\omega t+kx+n\theta)}}\\
\tilde{\u} &=\u-\uz\\
\fk        &= \inprod{-\tilde{\u}\cdot\nabla\tilde{\u}}{e^{i(\omega t+kx+n\theta)}}.
\end{align}

This finally yields equations for the fluctuations that are linear in $\uk$, and a base flow equation,
\begin{align}
i\omega\uk &= \Lk \uk + \fk, \quad \forall (k,n,\omega)\neq(0,0,0),\label{eqn:linu}\\
0 &= \fz -\uz\cdot\nabla\uz + \frac{1}{Re}\nabla^2\uz.
\end{align}

The unknown constant forcing $\fz$ describes the maintenance of $\uz$ via the radial derivative of the Reynolds stresses, generated from interaction with the other modes. We can identify $\uz$ with the turbulent mean velocity profile. Similarly, $\fk$ describes the excitation of $\uk$ by the triadic interaction with other wavenumbers. We cannot solve these equations without additional information because $\Lk$ incorporates $\uz$ and $\fz$ incorporates $\tilde{\u}$. This is essentially an appearance of the closure problem. However, the reader will notice that the perturbation equation (\ref{eqn:linu}) is a linear system with an unknown forcing $\fk$. This fact is central to our treatment.

We avoid the closure problem for the base flow equation simply by knowing \emph{a priori} the mean profile from experimental data. This allows calculation of $\Lk$ and precludes the need for an eddy viscosity formulation normally required, e.g. by \citep{Reynolds72}\cite{delAlamo06}, etc.

While $\fz$ can be simply calculated from the mean profile, we do not know the Reynolds stress at any other wavenumber combination, $\f_{kn\omega}$. Our approach is not to make assumptions about this forcing, but to simply examine the response of (\ref{eqn:linu}) at individual $(k,n,\omega)$ triplets over the set of all possible harmonic forcings.

The linear operator has the explicit form for pipe flow
\begin{equation}
\Lk=\left(
  \begin{array}{c}
    {Re^{-1}}(\partial_r^2 + {r^{-1}}\partial_r-n^2 r^{-2}-k^2-r^{-2})-iku_{000} ~~~~~~~~~~ -2inr^{-2} ~~~~~~~~~~~~~~~~~~ 0 \\
    2inr^{-2} ~~~~~~~~~~~~~~~~~~~~ {Re^{-1}}(\partial_r^2 + r^{-1}\partial_r-n^2r^{-2}-k^2-r^{-2})-iku_{000} ~~~~~~~~~~ 0 \\
    -\partial_ru_{000} ~~~~~~~~~~~~~~~~~~~~~~~~~~~~~~~~~~~~~~~~ 0 ~~~~~~~~~~ {Re^{-1}}(\partial_r^2 + r^{-1}\partial_r-n^2r^{-2}-k^2)-iku_{000} \\
  \end{array}
\right),
\label{eqn:Lpipe}
\end{equation}
where, as before, the states in $\bf{u}$ are the radial, azimuthal and axial velocities respectively.

We make the following observations:
\vspace{0.1in}
\begin{enumerate}
\item Only integer $n$ are permissible;
\item We expect downstream travelling waves such that $k$ and
$\omega$ are of opposite sign: henceforth $k > 0$ and $\omega \rightarrow -\omega$;
\end{enumerate}
\vspace{0.1in}

While Equation~\ref{eqn:linu} is linear in $\uk$, and $\Lk$ is identical to the operator obtained by linearising around the turbulent mean velocity profile, no linearisation has been performed. Nonlinear effects at other wavenumber-frequency combinations are retained through the action of the forcing $\fk$. In physical terms, this forcing can be considered to stimulate fluctuations that may lead to a net energy gain because of the characteristics of $\Lk$.

The model fully describes the energetic interaction between the base flow and fluctuation $\uk$, given a mean profile. Additionally, $\fk$ acts perpendicular to $\uk$, that is, $\inprod{\fk}{\uk}=0$. This can be interpreted as $\fk$ being conservative with respect to fluctuation energy, and so responsible for the transfer of energy in spectral space but not directly responsible for the extraction of energy from the base flow. A similar formulation to the current one, using this fact to derive a globally laminarising control law, was described in~\cite{Sharma06AIAA}.

Next, we proceed to analyse the response of $\Lk$ at a particular wavenumber combination, subjected to the harmonic forcing $\f_{kn\omega}$.

\subsection{Resolvent norms and model formulation}
Equation~\ref{eqn:linu} can be rearranged as
\begin{equation}
\uk = (i\omega I - \Lk)^{-1} \fk.
\label{eqn:model}
\end{equation}
The operator $(i\omega I - \Lk)^{-1}$ is called the resolvent and is the focus of our analysis. It provides a measure of the turbulent energy response that is possible for a given forcing. Some interpretation of the resolvent is given in Appendix A.

For pipe flow, using (\ref{eqn:Lpipe}) and for $(k,n,\omega)\neq(0,0,0)$, the resolvent can be written as:
\begin{equation}
(i\omega I - \Lk)^{-1}=
\left(
  \begin{array}{c}
    -{A}{Re^{-1}}+ikU - i\omega ~~~~~~~~~~ B ~~~~~~~~~~~~~~~~~~ 0 \\
    -B ~~~~~~~~~~~~~~~~~~~~ -{A}{Re^{-1}}+ikU - i\omega ~~~~~~~~~~ 0 \\
    \partial_r U ~~~~~~~~~~~~~~~~~~~~~~~~~~~~~~~~~~~~~~~~ 0 ~~~~~~~~~~ -{A}{Re^{-1}}+ikU - i\omega \\
  \end{array}
\right)^{-1}
\label{eqn:resolvent}
\end{equation}
with $A = (\partial_r^2 + \frac{1}{r}\partial_r-n^2r^{-2}-k^2-r^{-2})$, $B = 2 in r^{-2}$ and $U=u_{000}$. Equation~\ref{eqn:resolvent} clearly highlights that the operator is not self-adjoint in the presence of $\partial_r U$. The shear is a source of non-normality under the energy norm, coupling the radial and axial velocity components. The $A/Re$ term will always be small for Reynolds number large enough for turbulent flow.
One might expect large values of the resolvent norm under any of the following conditions:
\begin{enumerate}
\item in regions of high shear, where the $\partial_rU$ is large;
\item at critical layers where $\omega/k = U(y)$, so the component of the normal speed of propagation of the wave in the streamwise direction is equal to the local mean velocity;
\item for stationary modes with $k=\omega=0$.
\end{enumerate}

\subsection{Most amplified modes}

We seek a decomposition of the resolvent at a particular wavenumber pair and frequency which ranks
the response to forcing in some sense. We take the Schmidt decomposition (called
the singular value decomposition in the discrete case) of the resolvent, namely
\begin{equation}
\label{decomposeG} (i\omega I - \Lk)^{-1} = \sum_{j=1}^{\infty}
\psi_{j}(k,n,y,\omega)\sigma_{j}(k,n,\omega)\phi^{*}_{j}(k,n,y,\omega)
\end{equation}
with an  orthogonality condition
\begin{eqnarray} \int_{y} {\phi_{l}(k,n,y,\omega)}{\phi_{m}(k,n,y,\omega)}
\ud y=\delta_{lm}\\ \int_{y} {\psi_{l}(k,n,y,\omega)}{\psi_{m}(k,n,y,\omega)} \ud y=\delta_{lm}
\end{eqnarray} 
and
\[\sigma_{l}\geq \sigma_{l+1} \geq 0.\]

The $\phi_{j}$ and $\psi_{j}$ form the right and left Schmidt bases for the forcing and velocity
fields and the real $\sigma_{j}$ are the singular values. This decomposition exists if there are no
eigenvalues of $\curly{L}$ with zero real part and is unique up to a pre-multiplying unitary complex factor
on both bases corresponding to a phase shift and up to
the ordering of the $\sigma_{j}$'s \citep{Young-Hilbert}.

This basis pair can then be used to decompose arbitrary forcing and the resulting velocity at a particular Fourier
component
\begin{eqnarray}
\fk = \sum_{l=1}^{\infty} \phi_{l\k}a_{l\k} \label{decompf}\\
\uk = \sum_{l=1}^{\infty} \sigma_{l\k}\psi_{l\k}a_{l\k}.
\label{decompv}\end{eqnarray}
The energy of the same Fourier component of the resulting disturbance velocity is
\begin{equation}
E_{\k} = \inprod{\uk}{\uk} = \sum_{l=1}^{\infty}
\sigma_{l\k}^{2}a_{l\k}^{2}.
\end{equation}

Clearly the forcing shape that gives the largest energy at a particular frequency and wavenumber is
given with $a_{l\k}=0,~{l \neq 1}$. This approach permits the investigation of the dependence of maximum
energy amplification on the form of the forcing in the wavenumber and frequency domain.  The singular
value decomposition for a given wavenumber pair and frequency corresponds to full volume, three component forcing
and response modes ranked by the receptivity of the linear Navier-Stokes operator. The velocity response must have the same $k$ and $n$ but not necessarily
the same $y$ distribution (spatial phase variation in $y$) as the forcing.

By Parseval's theorem, the energy integrated over frequency and wavenumber is equal to that
integrated over the temporal and spatial domains (the spectral and physical spaces are isomorphic). As
such, the $\curly{L}_2$ norm of the resolvent is its leading singular value, $\sigma_{1}$. This means that the
normalised harmonic forcing that gives the largest disturbance energy in the $\curly{L}_{2}(\Omega
\times [0,\infty))$ sense is $f=\phi_{1}$, with a `gain' of $\sigma_{1}$. The next largest arises from
$f=\phi_{2}$ and so on, at a particular wavenumber pair and frequency.  The corresponding flow
response modes are given by the related $v = \psi^*_1, \psi^*_2$, etc. For $\sigma_{j}$ near zero,
the modes are not easily computed because they are effectively degenerate. However for the leading
singular values the mode shapes are extremely robust to numerical error.

This decomposition permits analysis of what we call the forcing
and response modes associated with large responses of the flow (the ``optimal response'').
In this sense, this decomposition analyses the receptivity of the flow to forcing.

In what follows, we focus our attention on the modes associated with the first singular value for a range of wavenumber-frequency combinations $(k,n,\omega)$ and show that they agree very well with experimental observations and classical scaling concepts.

\subsection{Computational approach}

The computational analysis of the linear operator, $\curly{L}$, was performed using a modified version of the spectral code of \cite{Meseguer03}.
The code essentially provides the operator $\Lk$ described above, evaluated at a number $N$ of wall-normal grid points. For the particular problem under consideration here, the only modification to the linear operator used by \cite{Meseguer03} is the use of the turbulent mean velocity profile instead of the steady, parabolic laminar base flow, as discussed above. We retain the same non-dimensionalisation, namely using the centreline velocity for a laminar flow with an equivalent mass flux, the steady laminar flow pressure gradient and the pipe radius, $a$.  This formulation is equivalent to using the bulk-averaged velocity, ${U_{bulk}}$, and pipe diameter, $D$, as velocity and lengthscales, which is more natural to the turbulent problem.  Of course, because the turbulent velocity profile is blunter than a laminar one, the constant mass-flux constraint means that the non-dimensional turbulent centreline velocity will always be less than one.

The form of the turbulent mean profiles was determined directly from experimental data obtained using Pitot probes in the Princeton/ONR \emph{Superpipe} and reported by \cite{mckeonmean04}.  This experimental data spans the Reynolds number range $31 \times 10^3 \le Re \le 35 \times 10^6$ or $860\le R^+ \lesssim 535\times 10^3$, where $u_\tau=\sqrt{\tau_w/\rho}$ is the friction velocity, $\tau_w$ is the mean wall shear stress and $\rho$ is the density.  Interpolation between Reynolds numbers was performed in a process equivalent to assuming a Reynolds-number-independent form of the velocity profile. Issues of numerical stability limit our study to Reynolds numbers $Re\le 10 \times 10^6$, well below the estimate of the Reynolds number at which the Superpipe results may show some effect of wall roughness, $Re \sim 24 \times 10^6$ \citep{Shockling06}.  Note that this Reynolds number range also spans conditions that have been described as representative of \emph{high Reynolds number} turbulence in pipe flow by \cite{mckeoninertial07} ($R^+ \gtrsim 5\times 10^3$) and in boundary layers by \cite{HutchinsPTRSA07} ($\delta^+ \gtrsim 4\times 10^3$).

The very sensitivity of the operator to perturbations that leads to large energy amplification can also cause numerical stability issues associated with the spatial resolution employed in the pipe cross-section. Judicious choice of this resolution, $N$, is required. This is briefly discussed in Appendix B.

\subsection{A model of spectral energy distribution}

While the concept of the Navier-Stokes equations as a linear system with a non-linear feedback forcing was been raised in a
control theory context by \cite{Sharma06AIAA}, the emphasis here is more toward the receptivity of the flow, as explored for laminar pipe flow by \cite{Sharma09AIAA}. This approach is comparable to the ``mother-daughter'' scenario of \cite{Boberg88},
where a linear but non-normal process allows small disturbances to feed more energetic disturbances
that can dissipate the gained energy. Non-linear effects then transfer some of this energy to the
smaller initial disturbances. The understandings differ to the extent that the ``mother-daughter'' scenario considers the evolution of structural perturbations, and naturally leads to the transient growth problem of \cite{Butler92} and others. In contrast, the current work considers the gain response to harmonic forcing, naturally leading to a linear input-output analysis.

The modes under investigation are propagating in the streamwise and spanwise directions and distributed in the wall-normal direction. As such they are strongly analogous to the spectral decomposition of spatial and temporal velocity fields from experiments and simulations. Their wave-like nature implies that the propagation (phase) velocity of each mode is given by $U_w=\omega/K$, where $K=(k^2+n^2)^{1/2}$, with streamwise (normal) component $U_x = \omega/k$.   In global terms, we expect that this analysis will give some insight into the spectral energy storage. Under forcing of the appropriate magnitude at all wavenumbers and frequencies, the correct Reynolds stress tensor would be calculated, obtaining both the true variation of turbulent energy production and dissipation, and closure of the feedback to the given turbulent mean velocity profile.

In this spirit, the current work explores the $(k,n,\omega,Re)$ parameter space and validates the efficacy of our simple model, by demonstrating that it is capable of capturing classical and empirically-observed features of wall turbulence, such as inner and outer scaling, ``attached'' and ``detached'' motions and Reynolds number trends.

\section{Predictions of the model}

In this section, we describe the characteristics of the modes that are predicted by the preceding analysis. We consider only propagating, helical modes in the streamwise direction, with $k \ne 0$, $n\ne0$.

We begin by presenting results for Reynolds numbers $Re=75 \times 10^3$ and $Re = 410 \times 10^3$ ($R^+ = 1800$ and $8500$, respectively).  The former is representative of the upper range accessible with current DNS techniques, while the latter Reynolds number is high enough to ensure characteristics of ``high Reynolds number'' turbulence~\citep{mckeonmean04,mckeoninertial07,HutchinsPTRSA07}. In addition, we know the velocity profiles and the streamwise and wall normal turbulence statistics and spectra from the Superpipe at, or close to, these conditions. This facilitates comparison with experimental results \citep{mckeonmean04,Morrison04,Zhao07}. From these data, we expect that the streamwise turbulent energy across the pipe cross-section lies within the range $5 \times 10^{-2} \lesssim k \lesssim 5\times 10^2$.

\subsection{Perturbation mode shapes}

\begin{figure}
\begin{center}
\includegraphics[width = 14cm]{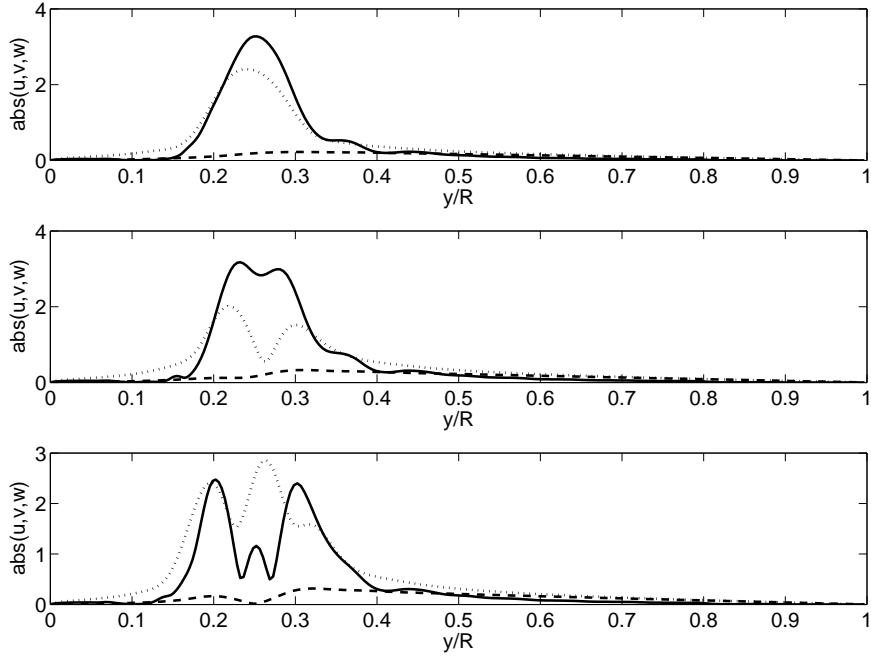}
\caption{Radial variation of the absolute amplitude of each component of velocity for the response modes with $(k,n,\omega)=(2,2,1)$ associated with the first three singular values, $\sigma_m$, at $Re= 75 \times 10^3$. ---: $abs(u)$; $-\,-$: $abs(v)$; $\cdots$: $abs(w)$.  Top, middle and bottom panels correspond to $m=1,2,3$, respectively.}
\label{fig:modeis}
\end{center}
\end{figure}

\begin{figure}
\begin{center}
\epsfxsize=13cm
\epsfbox{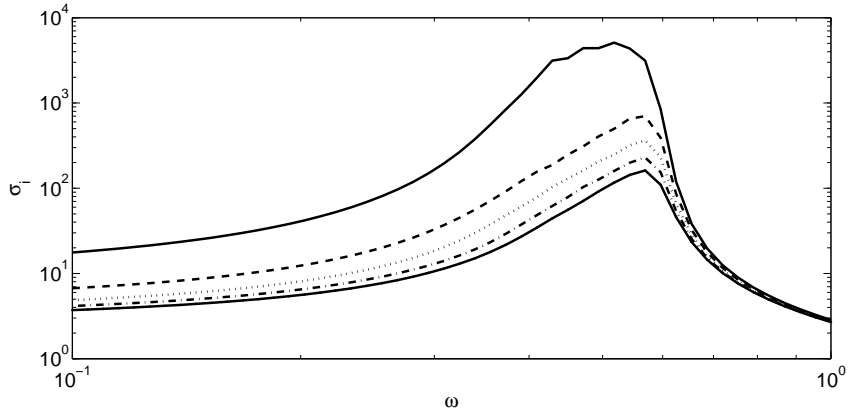}
\caption{Variation of singular values $\sigma_m$ with mode number, $m$, for $(k,n) = (1,10)$ and $Re=75 \times 10^3$. For a wide range of frequencies, the first singular value is an order of magnitude larger than the second.}
\label{fig:sigmavar}
\end{center}
\end{figure}

Figure \ref{fig:modeis} compares the velocity distributions for modes with the wavenumber pair $(k,n, \omega)=(2,2,1)$ for the first three singular values, $\sigma_m$ with $m=1,2,3$.  Clearly the higher order modes generate velocity distributions with more local maxima in the radial direction, apparently $m$ maxima in the streamwise velocity for the modes shown.
In addition, the singular value Bode plot of figure~\ref{fig:sigmavar} shows that the first singular value is significantly larger than the other singular values. This suggests that the mode corresponding to $\sigma_1$ is likely to dominate in observations in real pipe flow. By comparison, \cite{BaileyPOD09} have shown that more than 75\% of the streamwise energy is contained in the first radial POD mode for $Re = 150\times 10^3$. Thus we focus on the the first mode.

The literature associated with the amplification properties of the linearised Navier-Stokes operator indicates that maximum amplification of energy is obtained for the streamwise constant modes.  While we do not consider a $k$ of exactly zero in this study, we do observe similar trends as $k\rightarrow 0$. However we focus instead on the form of the velocity modes with streamwise wavenumber in the range corresponding to that observed in experiments.

\begin{figure}
\begin{center}
\includegraphics[width = 15cm]{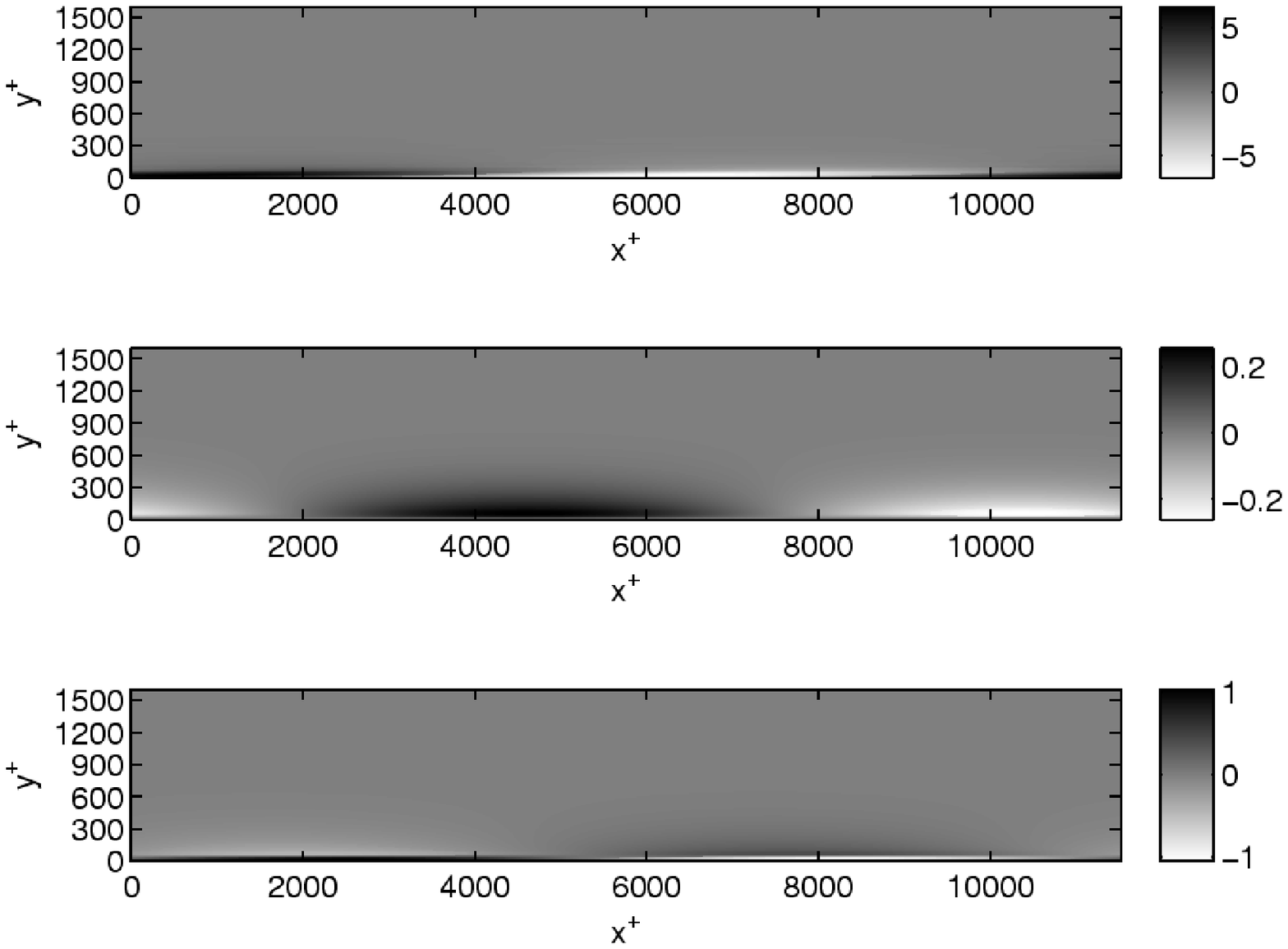}
\includegraphics[width = 7cm]{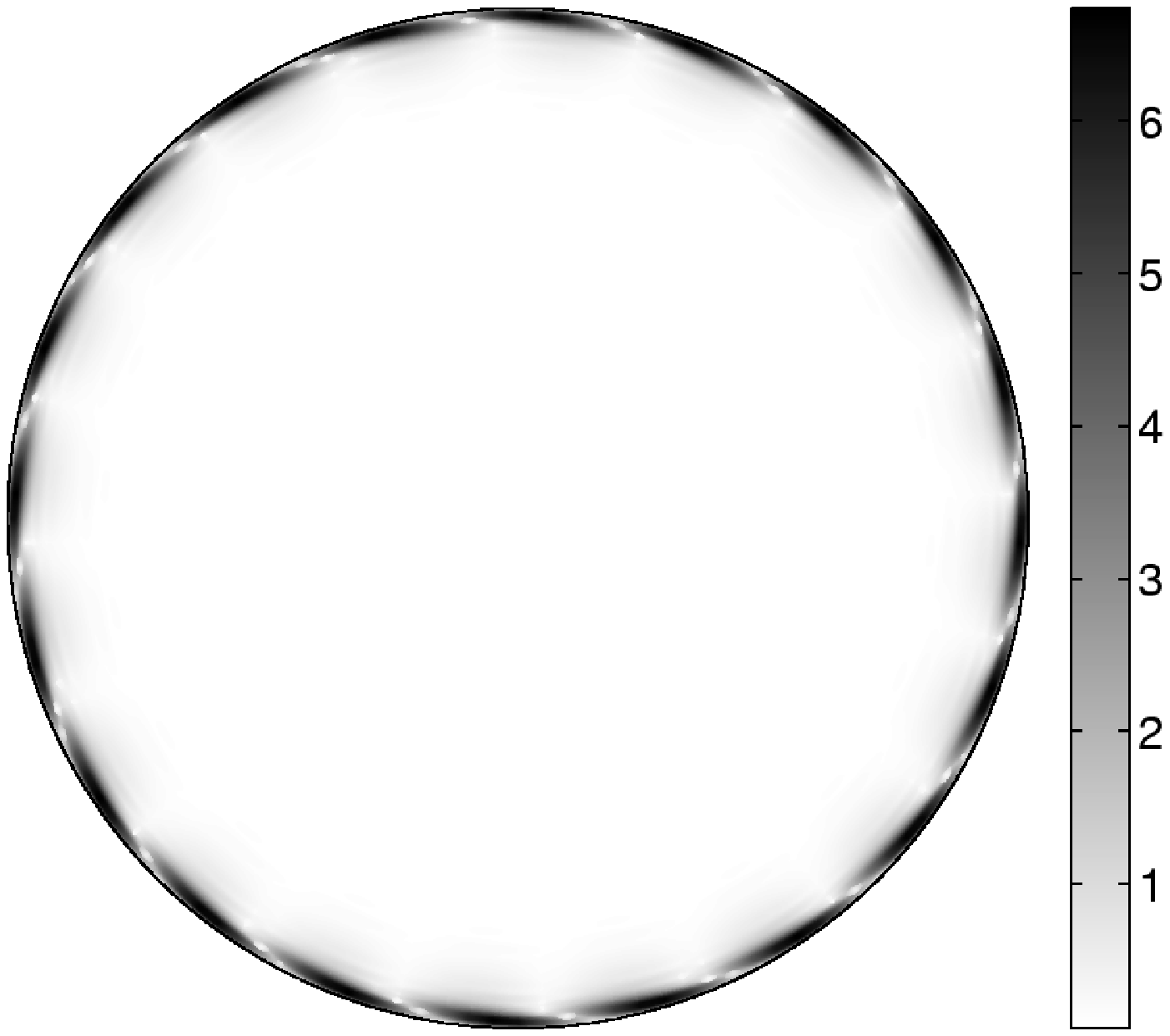}
\caption{Structure of the velocity field for the first mode with $(k,n,\omega) = (1,10,0.2)$ at $Re=75\times 10^3$. Top panels: $u,v$ and $w$ in the streamwise-wall-normal plane.  Bottom: variation of modal kinetic energy.}
\label{fig:modes1}
\end{center}
\end{figure}

\begin{figure}
\begin{center}
\includegraphics[width = 15cm]{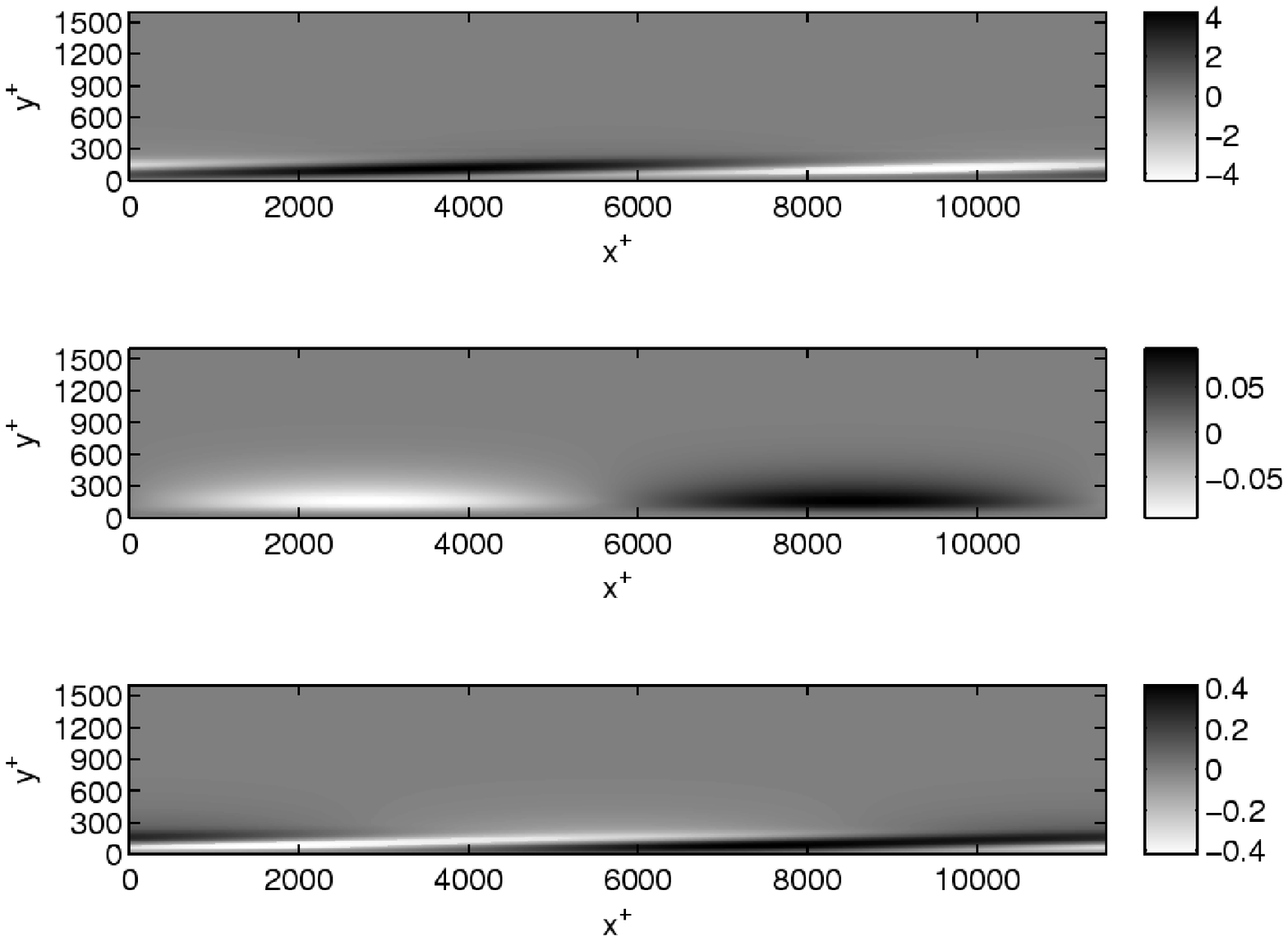}
\includegraphics[width = 7cm]{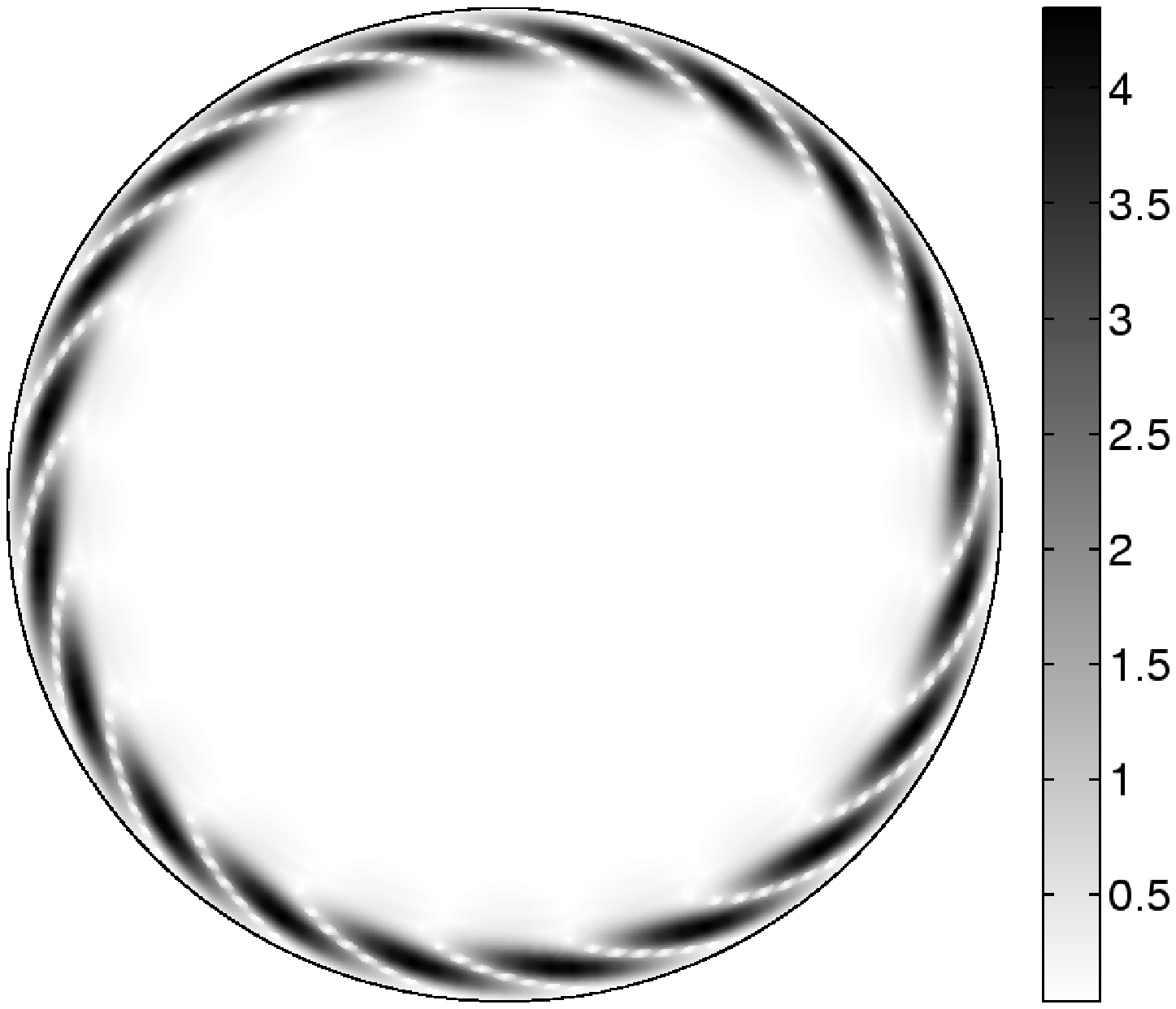}
\caption{Structure of the velocity field for the first mode with $(k,n,\omega) = (1,10,0.4)$ at $Re=75\times 10^3$. Top panels: $u,v$ and $w$ in the streamwise-wall-normal plane.  Bottom: variation of modal kinetic energy.}
\label{fig:modes2}
\end{center}
\end{figure}

\begin{figure}
\begin{center}
\includegraphics[width = 15cm]{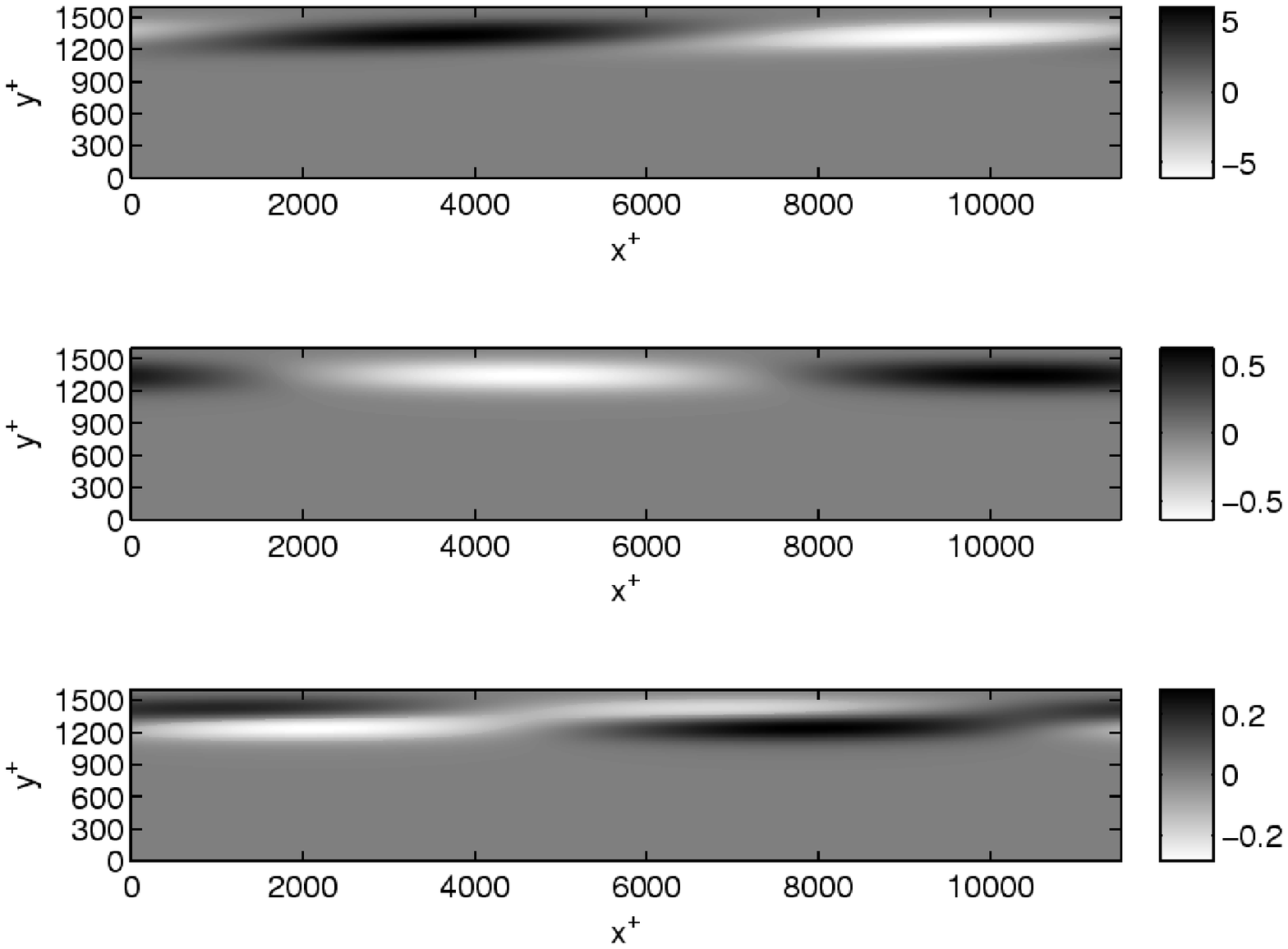}
\includegraphics[width = 7cm]{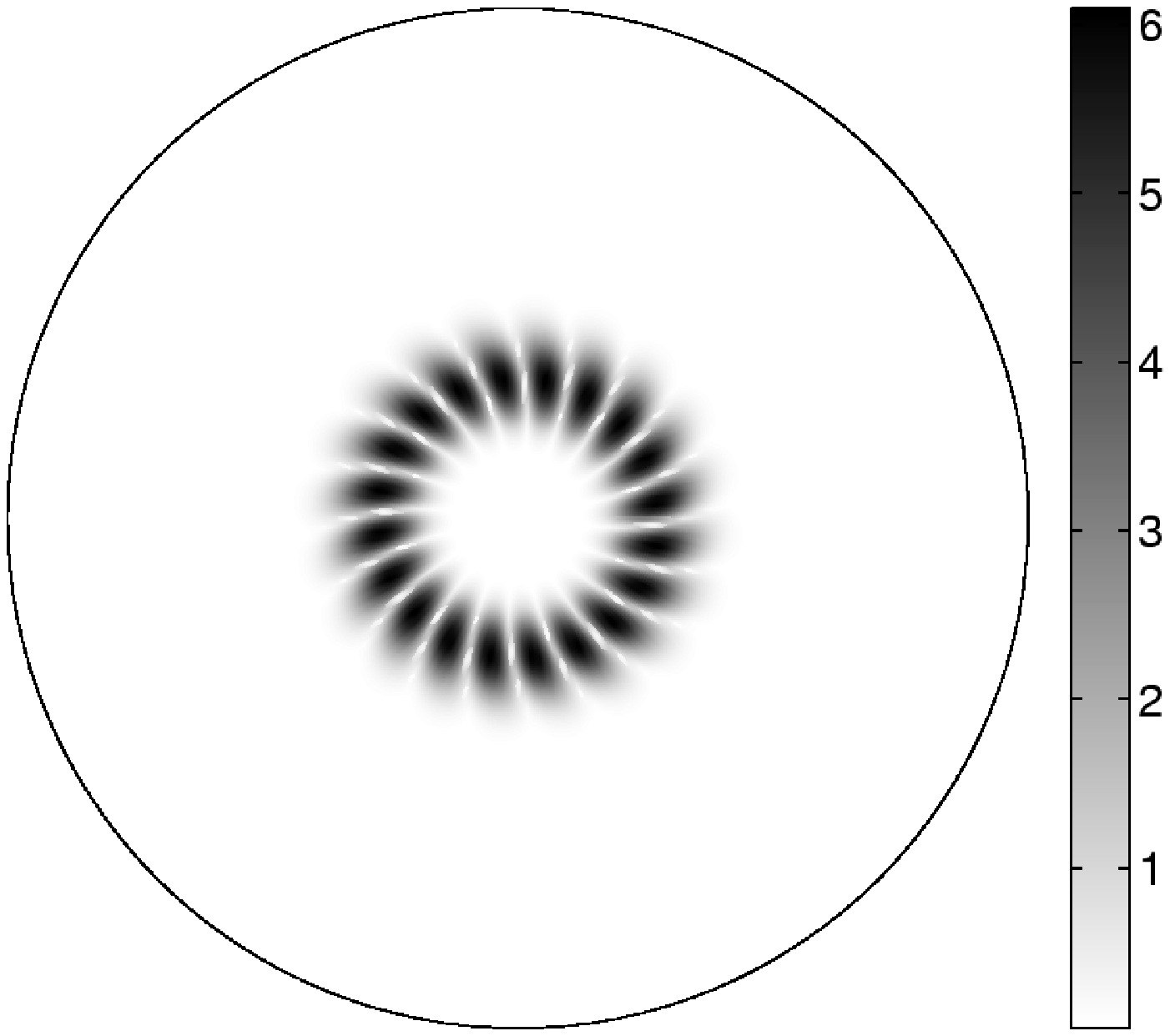}
\caption{Structure of the velocity field for the first mode with $(k,n,\omega) = (1,10,0.6)$ at $Re=75\times 10^3$. Top panels: $u,v$ and $w$ in the streamwise-wall-normal plane.  Bottom: variation of modal kinetic energy.}
\label{fig:modes3}
\end{center}
\end{figure}

Figures~\ref{fig:modes1}--\ref{fig:modes3} show the distribution of turbulent energy in the pipe for the first mode with wavenumber pair $(k,n)=(1,10)$ at $Re=75 \times 10^3$, an arrangement which will be shown later to be representative of a VLSM. The normal wave speed, $U_{w} = \omega/(k^2+n^2)^{1/2}$, is increased from $0.2$ to $0.6$ from figure~\ref{fig:modes1} to figure~\ref{fig:modes3}. Note that because $(k,n)$ is held constant between the figures the wavespeed in the streamwise direction, $U_{x}$, is a constant fraction of the normal wavespeed, and that the choice of $k\ne 0$, $n\ne0$ leads to helical mode shapes. As expected, the energy for the lower wave speed is concentrated close to the wall, in the inner (wall) region, while for the faster wave the energy is centred in the core of the pipe. This trend is to be expected from Taylor's frozen turbulence hypothesis if propagating waves are a realistic feature of wall turbulence. 

For this wavenumber pair, the forcing elicits an energetic response that is concentrated in the streamwise component for each $\omega$, with the relative magnitudes of the azimuthal and radial velocities being frequency-dependent.  The modes for the lower two frequencies shown in figures \ref{fig:modes1} and \ref{fig:modes2} have footprints which reach down to the wall. Increasing $\omega$ leads to a ``lifting'' of the streamwise and spanwise velocity components, manifested as a distinctive inclination to the wall of velocity isocontours.  At the highest of the three frequencies, shown in figure \ref{fig:modes3}, the mode detaches from the wall and the streamwise and spanwise velocities display something more like a two-level structure than inclined isocontours. For all three frequencies, the wall-normal component shows little phase variation with  wall-normal distance. The shapes imply a structure of long rolls with streamwise vorticity and strong streaks in the streamwise velocity. 
Note that the velocity distributions shown in figure~\ref{fig:modes2} are in excellent agreement with the conditionally-averaged very large scale structures determined 
for channel flow by \cite{Chung09} and the streamwise coherence inferred in from two-point correlations with a reference point outside the immediate near wall region in 
a high Reynolds number boundary layer by \cite{Guala09mod}, suggesting that modes of this kind may have particular importance for wall turbulence.
Section \ref{sect:inout} investigates modes with self-similar energy distributions in $y^+$ and $y/R$ for various wavenumber pairs. We call these ``inner'' and ``outer'' scaling modes respectively.

\subsection{Effect of streamwise wavespeed on the radial distribution of perturbation energy}

\begin{figure}
\begin{center}
\includegraphics[width=8cm]{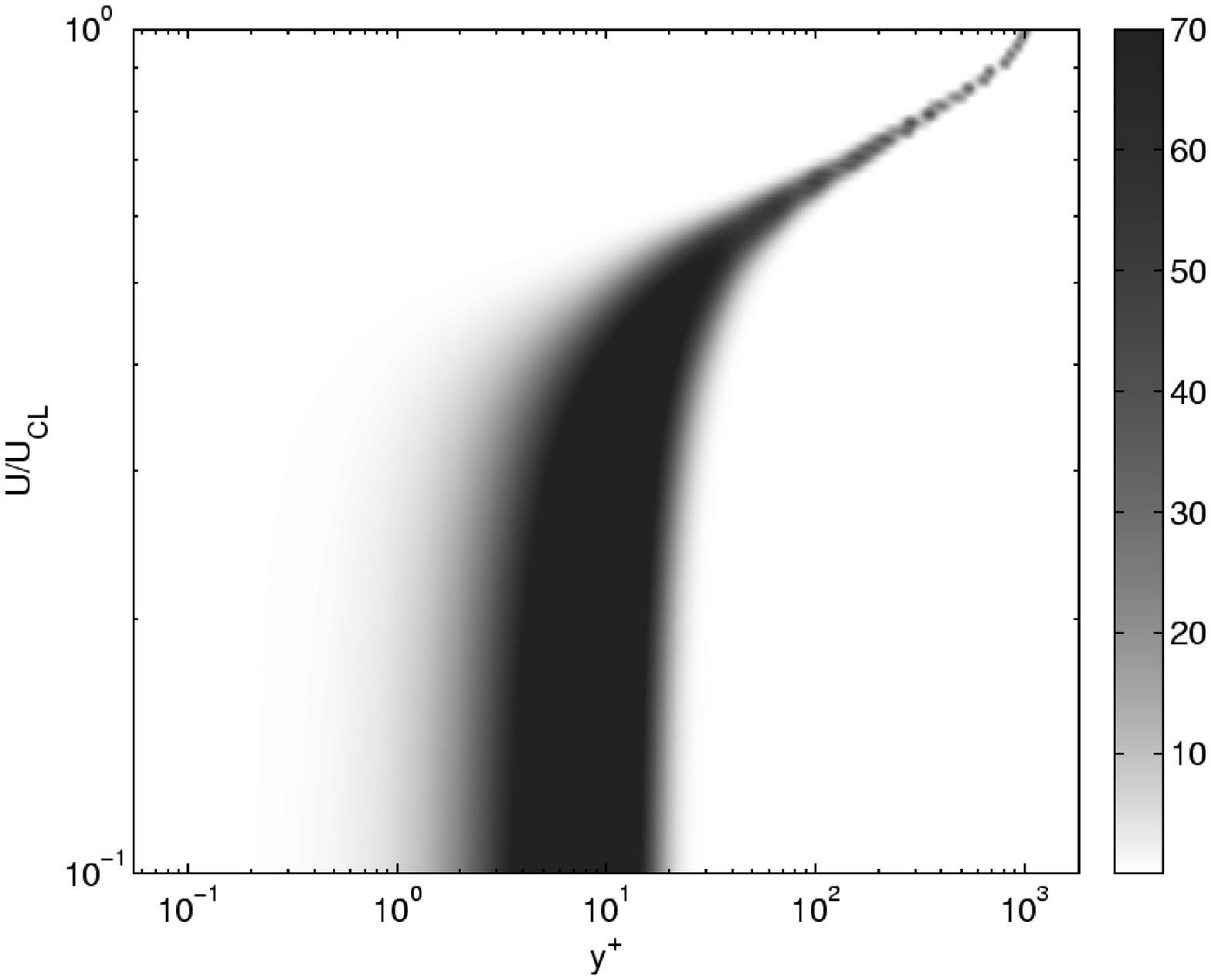}
\includegraphics[width=8cm]{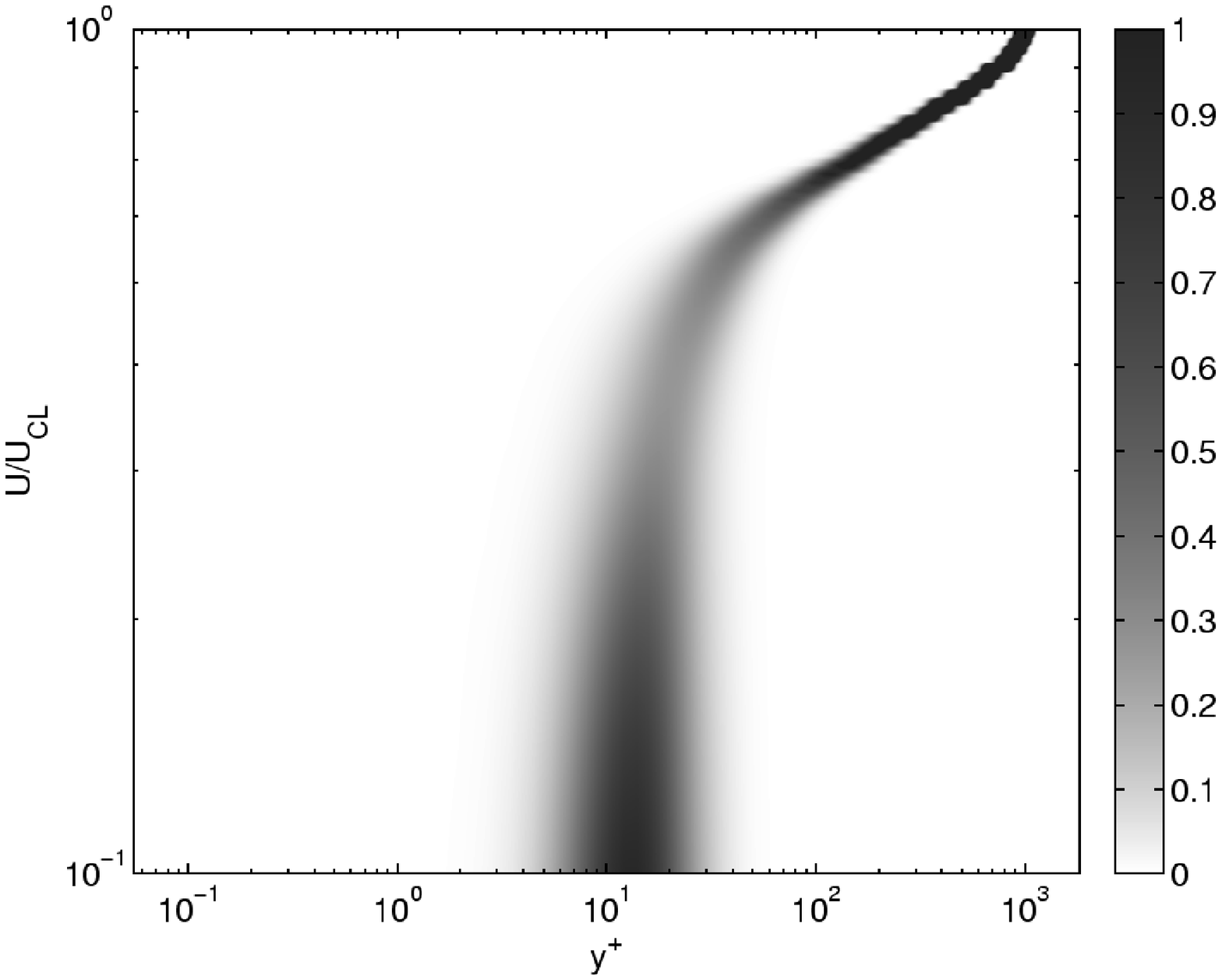}
\includegraphics[width=8cm]{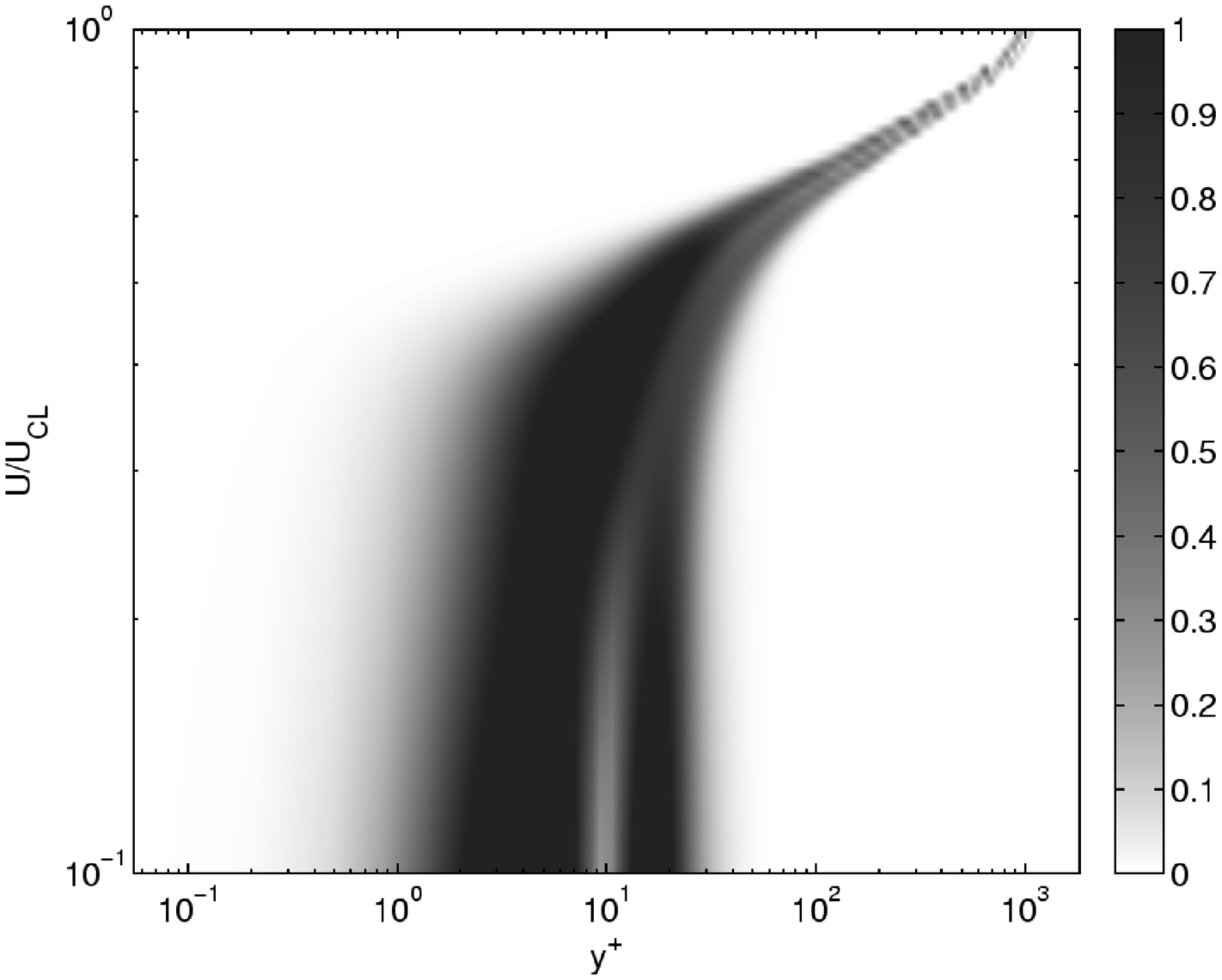}
\caption{Variation of energy distribution with phase speed for $\lambda_x^+ = 1000$, $\lambda_z^+ = 100$ at $Re=410 \times 10^3$ ($R^+ \sim 8500$). Top: $u$; middle: $v$; bottom: $w$.}
\label{fig:Re4p1e5_heat}
\end{center}
\end{figure}

In this section we explore the influence of frequency, or more accurately the streamwise wavespeed $U_{x}/U_{CL}$, on the mode shapes. The radial distribution of energy in each velocity component for the mode with $(k,n)=(2 \pi R^+/1000,2 \pi R^+/100)$ and streamwise wavespeed in the range $0.1 \le  U_{x}/U_{CL} \le 1$ at $Re=410 \times 10^3$ is shown in the composite contour plots of figure~\ref{fig:Re4p1e5_heat}. This wavenumber combination, $(\lambda_x^+,\lambda_z^+)\approx(1000,100)$ with an appropriate phase velocity $U_x^+ \sim 10$, appears to be representative of the signature of the near-wall cycle in many studies across a range of Reynolds numbers and flow types. In other words, it is a universal feature of wall turbulence responsible for the near-wall peak in the streamwise intensity and the wall shear stress. The changing amplitudes in each velocity component as the wavespeed increases reflect the distribution of energy between components and across the radius for different frequencies. The distributions are normalised to give an identical total kinetic energy for each mode.

The figure shows that for low wavespeeds the contours of constant energy lie at approximately the same wall-normal locations for all three velocity components and that the mode reaches down to the wall (i.e. is ``attached''). The distribution of energy between the velocity components is consistent with the well-known picture of streamwise rolls and streaks, in which the wall-normal and (double-peaked) azimuthal velocities have a similar order of magnitude and give rise to much larger fluctuating streamwise velocities.  This type of distribution is observed for this wavenumber combination at all Reynolds numbers considered, although high radial resolution is required to capture it accurately as the Reynolds number increases. This is directly related to the experimental and computational difficulties associated with an increasing range of scales.

A transition occurs in the radial and component-wise energy distribution with increasing wavespeed.  Instead of the near-wall roll/streak pattern described above, the peak energy moves away from the wall and the energy is concentrated in the core region. The wall-normal and streamwise energies are of similar order of magnitude (larger than the azimuthal component of energy), such that the mode pattern is more indicative of the signature of concentrated spanwise vorticity.

Aspects of the results for this wavenumber pair are representative of the trends observed for other modes. The location of the peak energy associated with a particular mode moves slowly away from the wall as the frequency is increased.
For a range of low $\omega$ ($U_x$), the energy is concentrated in the near-wall region. For higher $\omega$, the energy is concentrated in the core of the pipe.
The shape of the velocity profile near the wall constrains the range of possible mode shapes. Specifically, it constrains the radial extent of the mode, providing a minimum distance from the wall for the energetic peak in the streamwise stress. For example, in figure \ref{fig:Re4p1e5_heat}, this occurs at $y^+\simeq 10$. The wall-normal distance to the peak is Reynolds number dependent.

\subsection{Comparison of streamwise wavespeeds and local mean velocity}

\begin{figure}
\begin{center}
\epsfxsize=14cm
\epsfbox{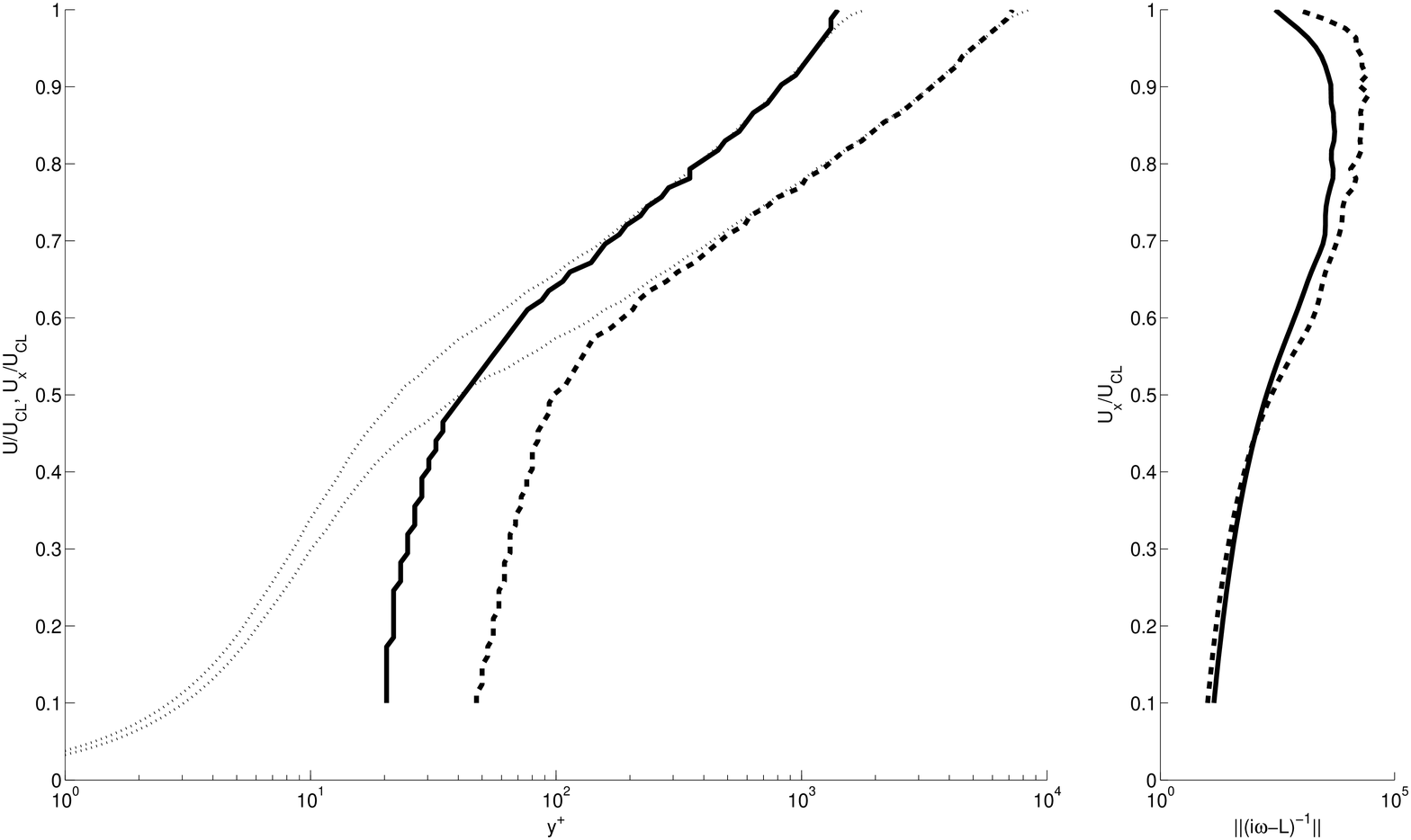}
\caption{Variation of location of peak streamwise perturbation amplitude with  streamwise phase speed for modes with $(k,n)=(1,10)$. ---: $Re=75\times 10^3$; $-\,-$ $Re=410\times 10^3$. The dashed gray lines show the turbulent mean velocity profiles at each Reynolds number. Right panels: variation of resolvent norm with streamwise phase speed.}
\label{fig:upV_Res}
\end{center}
\end{figure}

In the following discussion, we take the wall-normal location of the streamwise energy peak to be representative of a nominal mode centre. This allows comparison between the streamwise component of wavespeed and the local mean velocity. From figure~\ref{fig:Re4p1e5_heat}, we expect that these will not always be equal. Figure~\ref{fig:upV_Res} shows the variation of peak energy location with streamwise phase speed, for $(k,n)=(1,10)$ and Reynold numbers of $75\times 10^3$ and $410\times 10^3$. Whilst there is a region in the core of the pipe where the wave has a streamwise wavespeed that is similar to the local mean velocity at the peak of the streamwise energy, close to the wall there is a significant deviation between the two velocities. This implies that our assumption that the mode is centred at the streamwise energy peak is incorrect near the wall, that these modes do not obey Taylor's hypothesis , or that the modes will not be observed in the flow. In order to distinguish between the distinct behaviour near to and very far from the wall, we designate the former ``wall modes'' and the latter ``critical modes'', for reasons that will become clearer in section~\ref{sect:disc}. The right-hand panel of figure~\ref{fig:upV_Res} shows that the first singular value for this wavenumber pair increases with Reynolds number, at least in the region far from the wall.

\begin{figure}
\begin{center}
\epsfxsize=14cm
\epsfbox{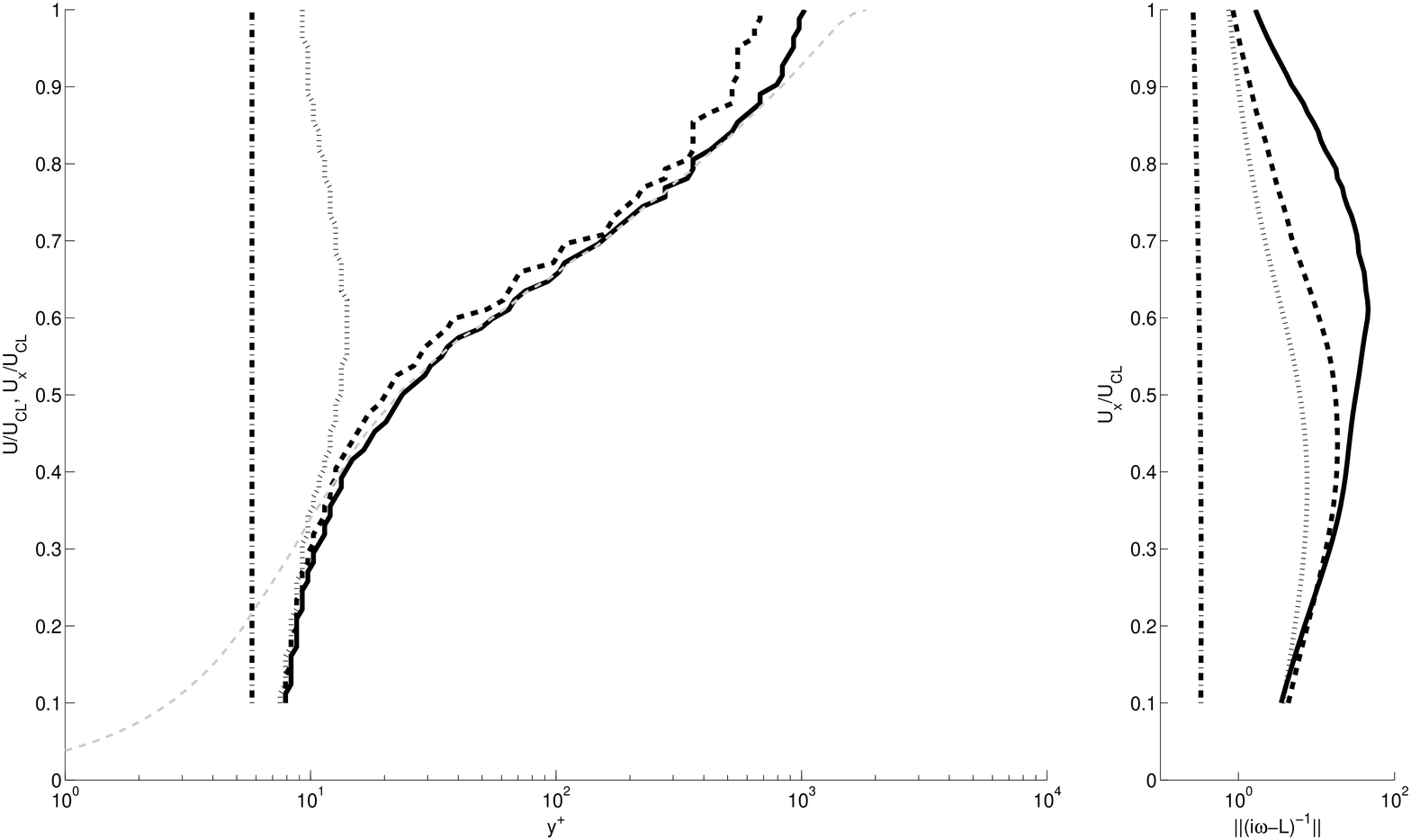}
\epsfxsize=14cm
\epsfbox{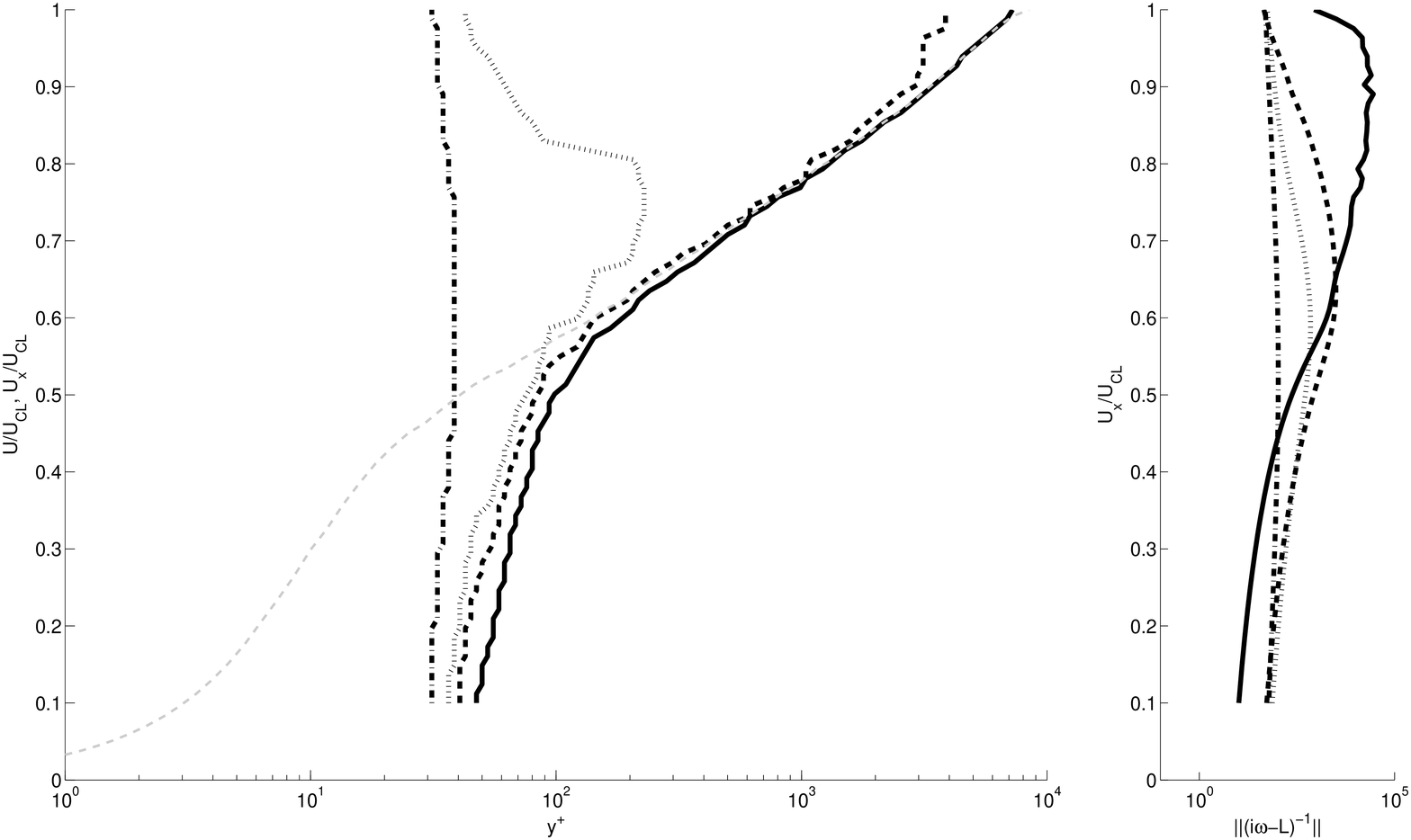}
\caption{Variation of location of mode peak amplitude with spanwise wavenumber. Top: $\lambda_x^+ \approx 1000$ and $\lambda_z^+=100$ (---), 50 ($- -$), 30 ($\cdots$) and 10 ($\cdot - \cdot$) at $Re=75 \times 10^3$. Bottom: $k=1$ and $n=10$ (---), 100 ($- -$), 200 ($\cdots$) and 400 ($\cdot - \cdot$) at $Re=410 \times 10^3$.  The dashed gray lines show the turbulent mean velocity profiles at each Reynolds number. Right panels: variation of resolvent norm with streamwise phase speed.}
\label{fig:phaseu_n}
\end{center}
\end{figure}

Figure~\ref{fig:phaseu_n} shows the influence of azimuthal wavenumber on the streamwise wavespeed corresponding to the peak in streamwise energy, for a mode representative of the near-wall cycle at $Re=75 \times 10^3$ and another mode with $(k,n)=(1,10)$ at $Re=410\times 10^3$, respectively. For azimuthal wavenumbers that are not ``too large'', both cases show critical modes in the core of the pipe over a $y^+$ range that increases with increasing $n$, with wall modes observed closer to the wall and a transition region in between.  However, when the azimuthal wavenumber becomes sufficiently large, only wall modes are observed. Then, the peak energy remains at an approximately constant wall-normal location independent of wavespeed and the mode shape is self-similar with decreasing wavespeed. In this case, the corresponding singular value is small and almost independent of streamwise wavespeed.

Wherever the streamwise wavespeed is equal to the prescribed local mean velocity, the behaviour approaches that of a ``critical'' layer.  Very large amplitude, spatially and temporally periodic waves have been observed in various shear flows \citep{Maslowe86} and explained in the context of critical layers.

Examination of the resolvent (Equation~\ref{eqn:resolvent}) shows that the system response approaches singular for a phase velocity close to the local mean $(\omega/k \sim U)$ and for high Reynolds number. That is, the flow approaches a state where it supports neutrally stable modes, corresponding to an eigenvalue approaching the imaginary axis.
Comparison between the right-hand panels of figure~\ref{fig:upV_Res} shows that the resolvent norm is orders of magnitude larger for such critical modes.
This is phenomenon is explored further in Section~\ref{sect:critlayer} below.

\subsection{Inner and outer scaling modes and the effect of Reynolds number}\label{sect:inout}

\begin{figure}
\begin{center}
\begin{minipage}{0.48\linewidth}
    \begin{center}
    \includegraphics[width = 7cm]{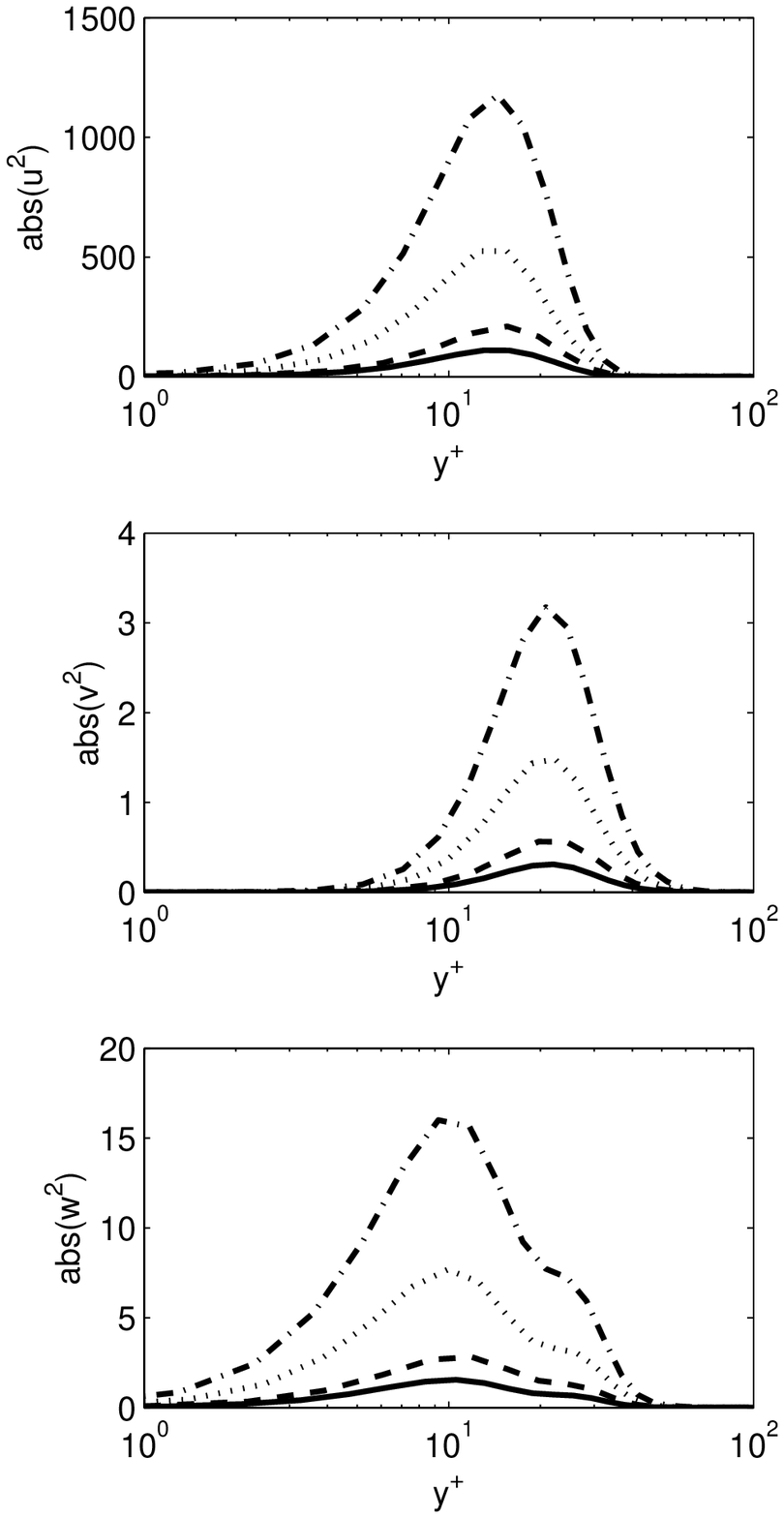}
    \end{center}
\end{minipage}
\begin{minipage}{0.48\linewidth}
    \begin{center}
    \includegraphics[width = 7cm]{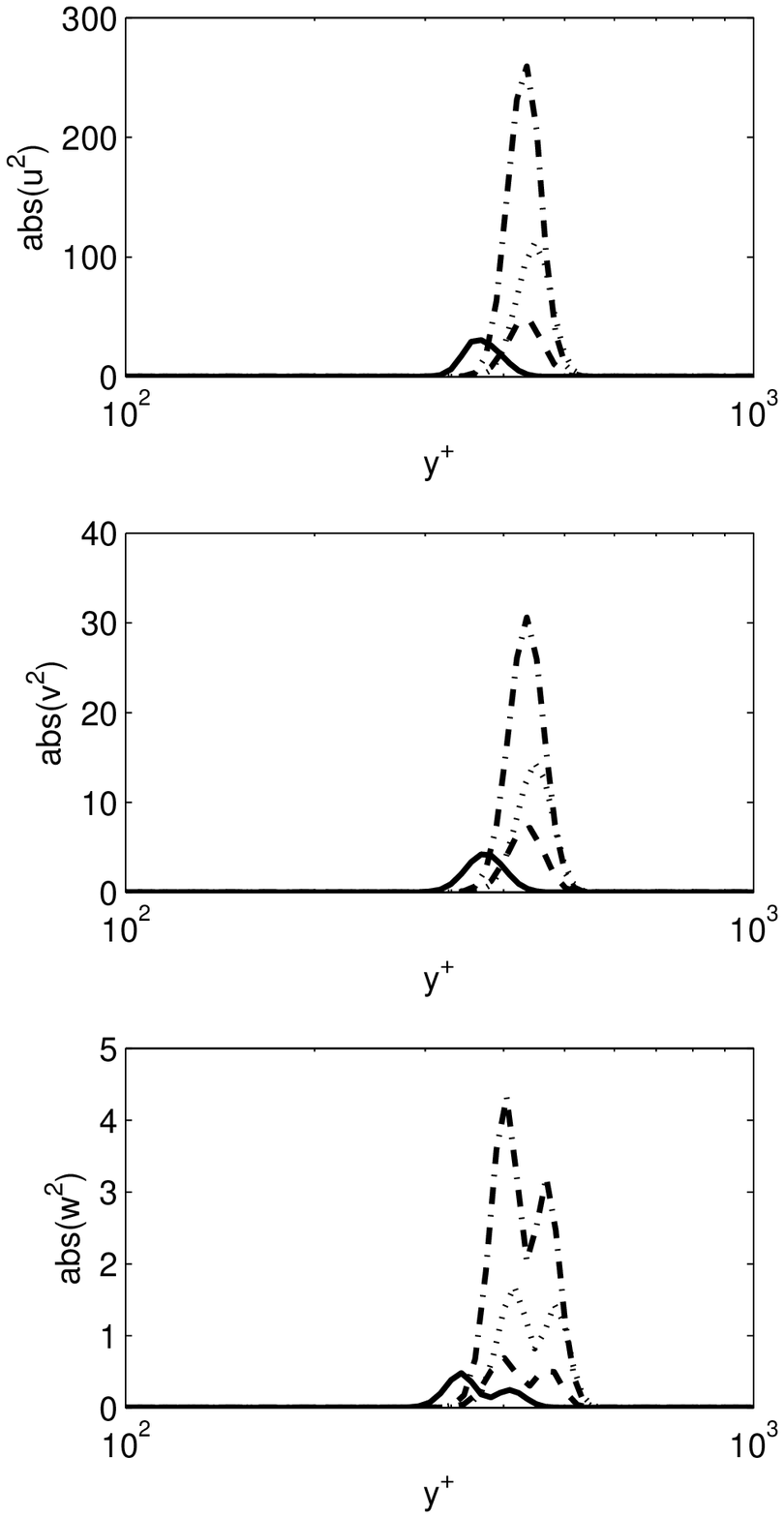}
    \end{center}
\end{minipage}
\caption{Energy distribution over the pipe radius for $(k,n)=(2 \pi R^+/1000,2 \pi R^+/100)$, or equivalently, $(\lambda_x^+,\lambda_z^+)\approx(1000,100)$. The equivalence is not exact due to the restriction of integer azimuthal wavenumber. The left panels show streamwise wavespeed $U_{x}^+ = 10$ and the right panels $U_{x}^+ = 20$. Reynolds numbers: --- $75 \times 10^3$, $- -$ $150\times10^3$, $\cdots$ $410\times10^3$, $\cdot - \cdot$ $1\times10^6$.}
\label{fig:lxpconst1}
\end{center}
\end{figure}

By carefully selecting the streamwise wavespeed, we observe scaling behaviour that is surprisingly consistent with that seen in experimental velocity spectra. The modes approach self-similarity when scaled on the classical inner or outer scales.  Figure~\ref{fig:lxpconst1} shows the $(u,v,w)$ mode shapes for a wavenumber-frequency combination that is representative of the near-wall cycle, $(k,n)\approx(2\pi R^+/1000,2 \pi R^+/100)$ with streamwise wavespeeds equal to $U_{x}^+ = 10$ and $20$. The former velocity is known to correspond to the convection velocity of near-wall structure, while the latter velocity will only correspond to the inner scaling region for sufficiently high Reynolds number, when the latter velocity occurs below the outer edge of the log region in the mean velocity.
It is understood that the velocity spectra at high wavenumbers near the wall collapse when scaled on inner units, while the lack of collapse of the intensities arises from the influence of larger scales \citep{Metzger01}.  Similarly, the intensities approach collapse in the core of the pipe as the Reynolds number is increased \citep{mckeoninertial07}.

Figure \ref{fig:lxpconst1} shows that these modes become self-similar when scaled in inner units across nearly two decades in Reynolds number and that they are ``attached'' in the sense that their footprints reach down to the wall.  Remarkably, the peak energy in the streamwise $u$ component for the lower wavespeed of $10 u_\tau$ is at the appropriate wall-normal location, $y^+ \approx 15-20$, obtained from experimental and computational observations of the near-wall cycle.  The increase in the maximum power occurs because the modes are co-located in plus units, so their extent decreases in dimensional units as the Reynolds number increases, and the total perturbation energy in each mode is normalised to one. There is an upper limit to the range of Reynolds number that can be considered due to numerical resolution. This is discussed in Appendix B. However the Reynolds number trend appears to be quite clear.

\begin{figure}
\begin{center}
\begin{minipage}{0.48\linewidth}
    \begin{center}
    \includegraphics[width = 7cm]{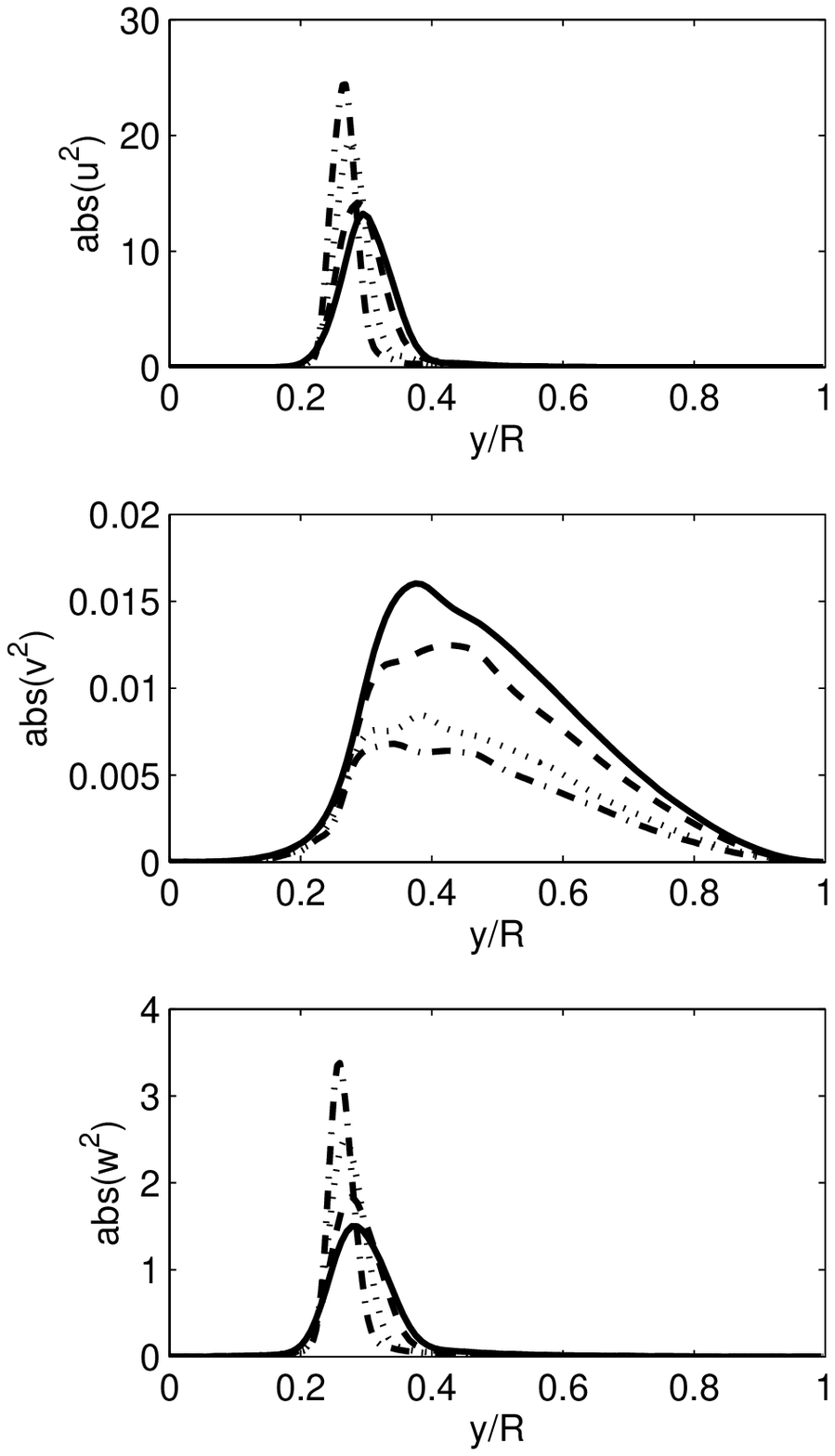}
    \end{center}
\end{minipage}
\begin{minipage}{0.48\linewidth}
    \begin{center}
    \includegraphics[width = 7cm]{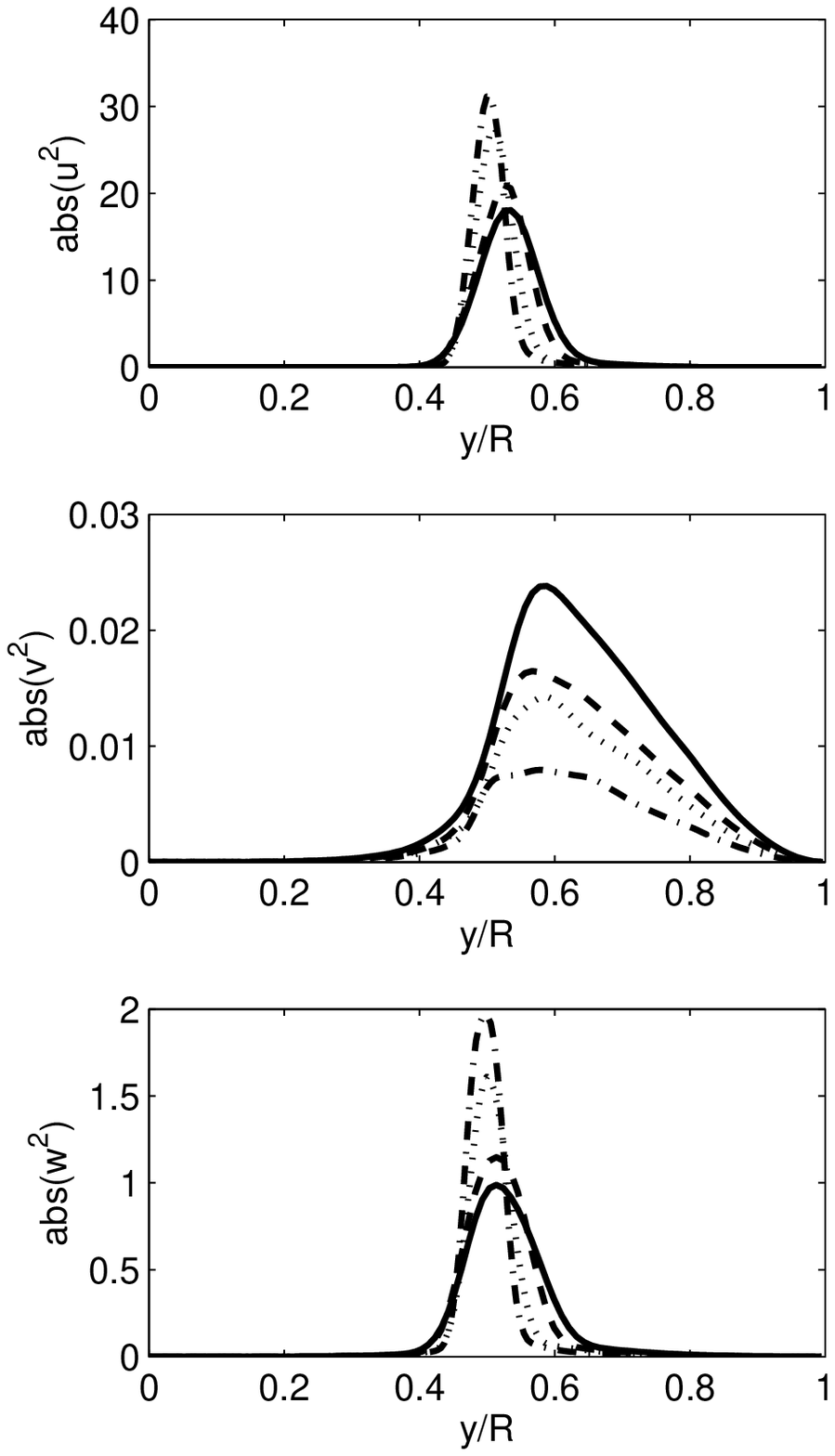}
    \end{center}
\end{minipage}
\caption{Energy distribution over the pipe radius for $(k,n)=(1,10)$. The left panel shows the distribution for streamwise wavespeeds such that the velocity deficit is $U_{CL}^+-U_x^+ = 4$ and the right panel for a velocity deficit of $U_{CL}^+-U_x^+ = 2$. Reynolds numbers: --- $75 \times 10^3$, $- -$ $150\times10^3$, $\cdots$ $410\times10^3$, $\cdot - \cdot$ $1\times10^6$.}
\label{fig:kconst1}
\end{center}
\end{figure}

A similar result can be obtained for modes that occur in the core of the pipe, i.e. they are expected to scale on outer variables. Figure~\ref{fig:kconst1} shows the $(u,v,w)$ mode shapes for $(k,n)=(1,10)$ and constant velocity defects, $U_{CL}^+-U_x^+$. The approach to similarity with increasing Reynolds number is slow, in agreement with the scaling of the integrated turbulence intensities \citep{mckeoninertial07}, with the wall-normal component collapsing earliest.

We see that a straightforward analysis of the resolvent, a linear operator, produces mode shapes that agree with known scaling remarkably well. In one respect this simply reflects the self-similarity of the mean velocity profile near the wall, in the sense that the modes are a result of the local shape of the velocity profile, which does not vary in inner units once the Reynolds number is sufficiently high.  However the recovery of this result underlines the utility of the model for describing turbulent flow. A more formal justification for the scaling will be given in the discussion presented in Section \ref{sect:disc}.

As evidenced by the failure of inner scaling to collapse experimental observations of the integrated streamwise energy near the wall, the transition from inner to outer scaling is complex and the range of modes exhibiting inner and outer scaling at a given flow condition is a function of the Reynolds number.  We explore this regime further in the next section.

\subsection{The very large scale motions and transition from inner to outer scaling}\label{sect:VLSM}

The existence of inner and outer scaling modes in our model is consistent with the classical turbulence scaling picture, in which viscous and outer scales are sufficient to describe the turbulent behaviour in the inner and outer regions of the flow.
However, as described above, recent work suggests that there is some outer influence on the inner structure. For instance, \cite{Metzger01,HutchinsPTRSA07} show that the position of the near-wall peak in streamwise energy weakly depends on Reynolds number.
We consider here a possible origin for this effect, focusing on the transition between inner and outer scaling of the velocity modes.

A concatenation of experimental results in wall turbulence in general, and specifically in pipe flow, have shown that a streamwise wavenumber $k=1$ is representative of the VLSM phenomenon \citep{Kim99,Morrison04,mckeoninertial07,Hutchins07,Monty07} although the recent study of \cite{Monty09} suggests that the exact details vary from flow to flow. The appropriate spanwise wavenumber is less clear; the spanwise measurements in pipe flow of \cite{Monty07} suggest $\lambda_z \approx 0.5R$ ($n \sim 12$), while the POD analysis and correlations of \cite{Bailey08} indicate that lower values, $n \sim 3$, are more realistic for the VLSMs. We investigate $n=10$ as a compromise that follows the expected critical layer aspect ratio discussed in section~\ref{sect:disc} and is in reasonable agreement with the work of both Monty and Bailey.

\begin{figure}
\centering
\includegraphics[width=9cm]{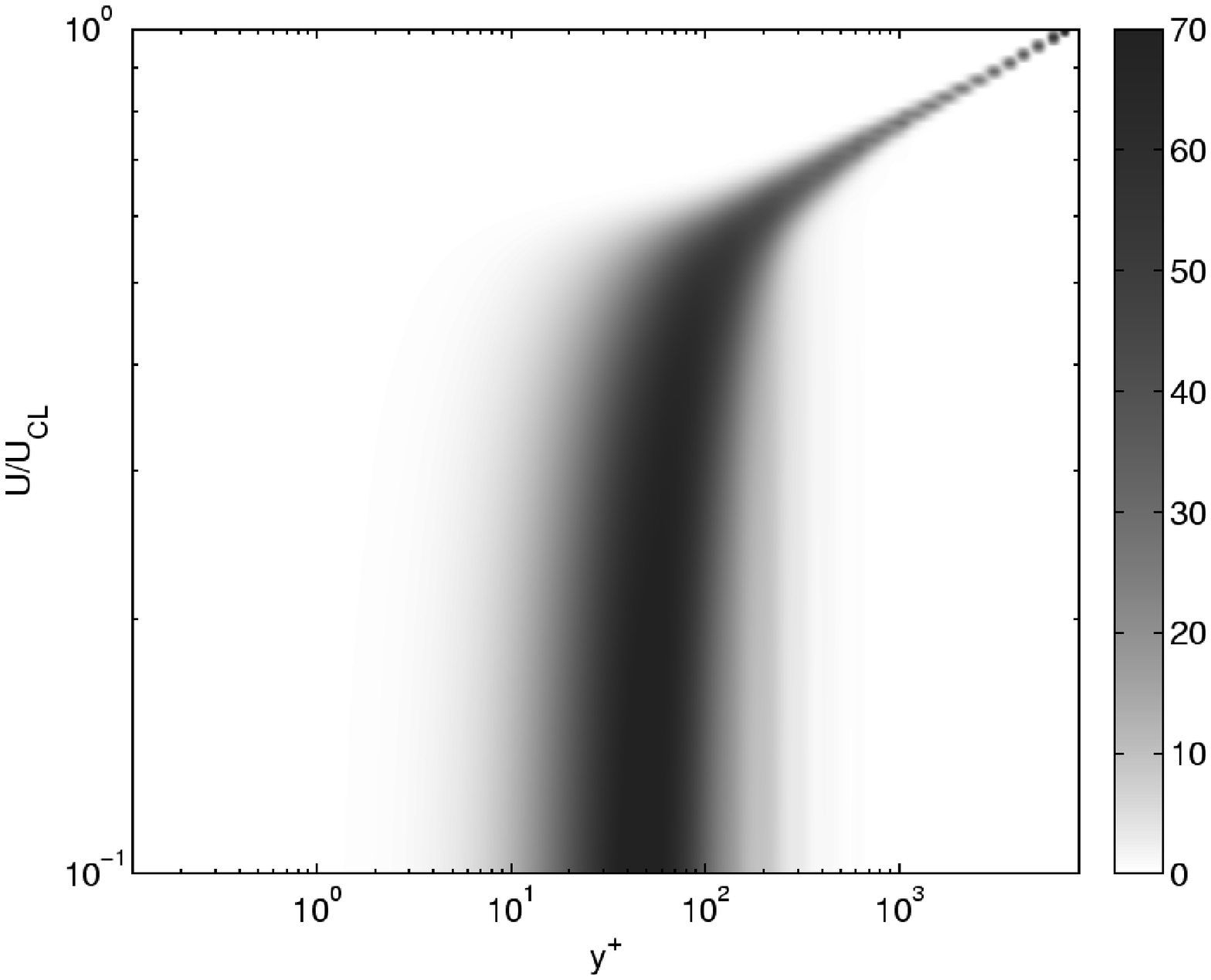}
\includegraphics[width=9cm]{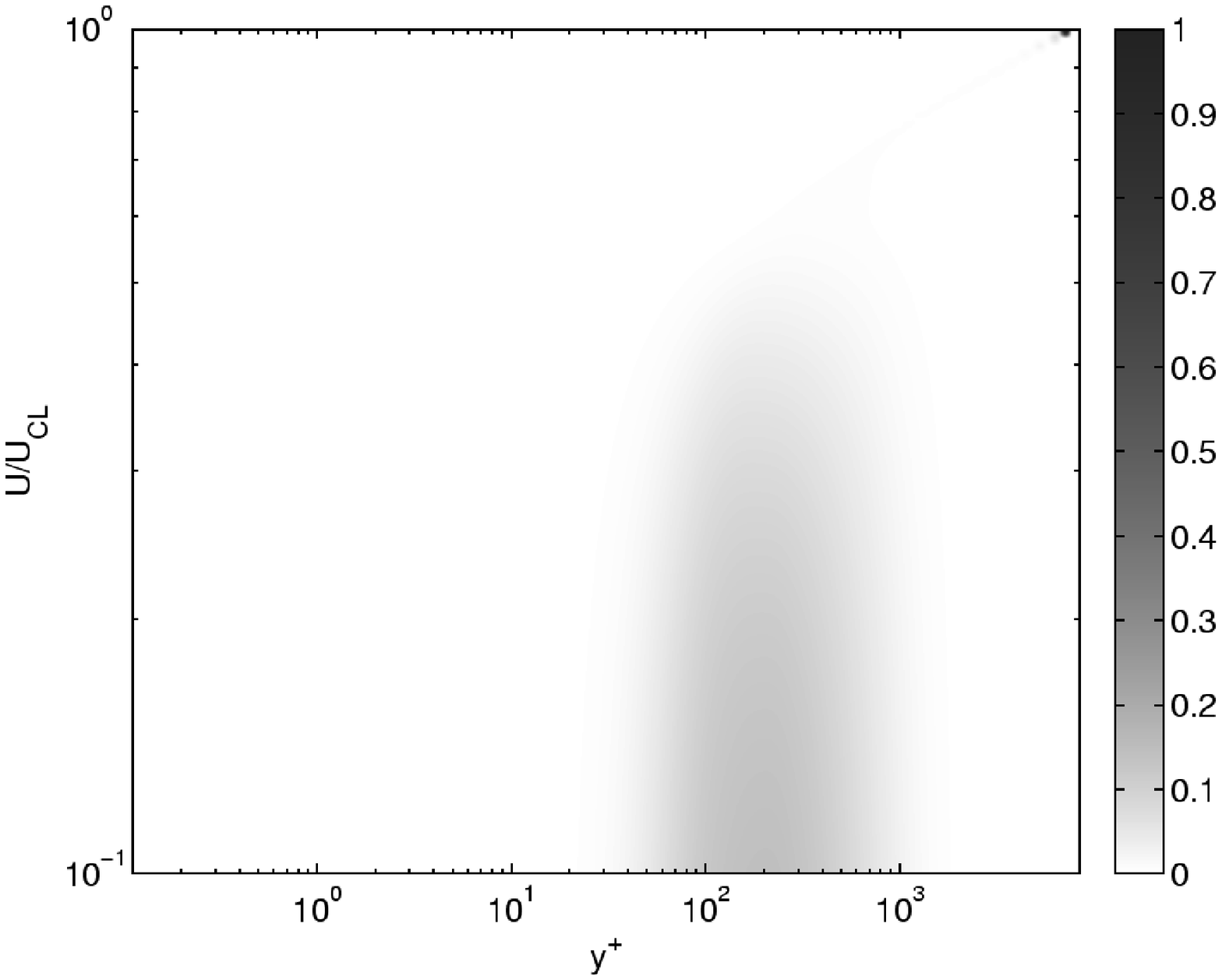}
\includegraphics[width=9cm]{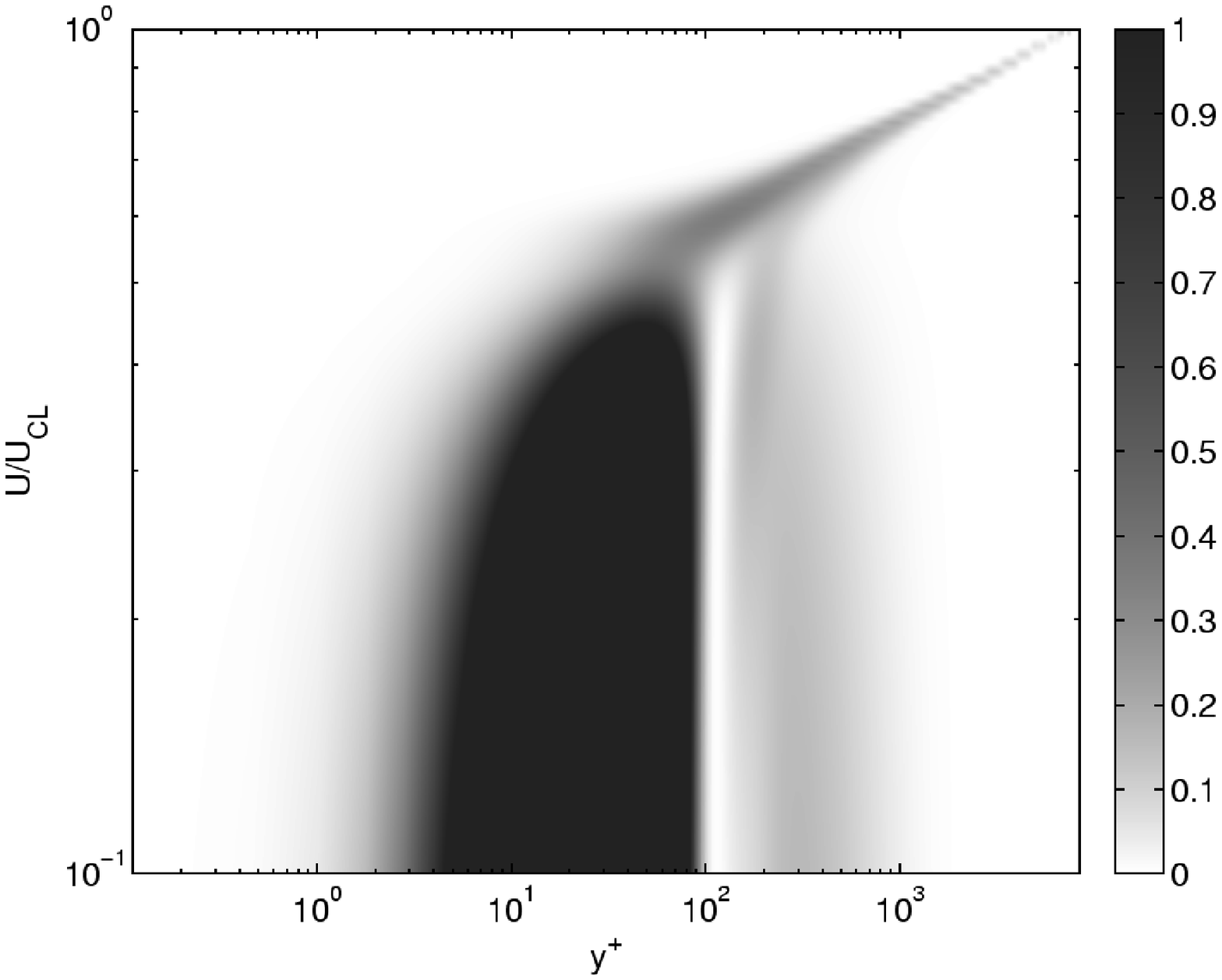}
\caption{Effect of phase speed on energy distribution for $(k,n)=(1,10)$ at $Re=410 \times 10^3$ ($R^+ \sim 8500$). Top: $u$; middle: $v$; bottom: $w$.}
\label{fig:V1}
\end{figure}

The variation of mode shape for $(k,n)=(1,10)$ with increasing phase speed is shown in figure~\ref{fig:V1} for $Re=410 \times 10^3$.  We have defined a mode as \emph{critical} when the streamwise wavespeed reaches the local mean velocity at the peak modal energy. The variation of this position with Reynolds number for the wavenumber pair related to the VLSM is illustrated in figure~\ref{fig:upV_Res}. In this case the mode first becomes critical when $U_{x}$ is a constant fraction of the centreline velocity, namely $U_{x}/U_{CL} \approx 2/3$, independent of Reynolds number. This relationship suggests that for the critical modes the appropriate scaling velocity is the centreline velocity, rather than the friction velocity.  This was also proposed by, among others, \cite{Jimenezlargescale04}, who observed that the so-called ``global'' modes in channel flow simulations appeared to convect with a velocity equal to $0.5 U_{CL}$.

\vspace{0.2in}

The foregoing section has demonstrated that the simple model for pipe flow is capable of reproducing several features of wall turbulence, from a self-similar distribution of energy for small scales near the wall to a large scale modal shape that is reminiscent of recently-observed conditional averaged structure in channels \citep{Chung09}, two-point correlations in medium and high Reynolds number boundary layers \citep{Guala09mod}, and energetic POD modes in pipe flow \citep{Duggleby07}. In a broad sense, the singular value decomposition selects the mode shape at each $(k,n,\omega)$ combination that is ``most likely'' to be observed based on receptivity to forcing at that combination, or ``optimal'' in a sense analogous to that of earlier initial value studies. The amplitude of the response is set by the amplitude of the forcing, which is determined by other $(k,n,\omega)$ combinations such that the ensemble of mode shapes, and specifically the Reynolds stress distribution, is consistent with the base mean flow. Note that a canonical spectral analysis records the integral energy over all modes with the same wavenumbers, so there is not a one-to-one relationship between the information given by the different bases corresponding to the power spectrum and the modal decomposition, except in an integral sense.

\section{A wall and critical layer framework}\label{sect:disc}

In this section, we show that the simple model presented in Section~\ref{sect:approach} predicts the location of the peak in streamwise turbulent energy associated with a wavenumber pair representative of a VLSM-type motion. We then offer a critical-layer interpretation of this type of mode.

\subsection{Scaling of the VLSM energy peak}\label{sect:VLSMpredict}

The peak energy in the streamwise velocity component for the first critical mode with $(k,n)=(1,10)$ occurs for $y/R < 0.1$ for all the Reynolds numbers considered here. Thus it is clear that the first critical mode occurs within the log region, at least for sufficiently high Reynolds numbers.\footnote{A logarithmic profile will give a reasonable approximation for the local velocity at the lower Reynolds numbers, even if the profile does not correspond to an overlap region that is independent of Reynolds number.} Based on this velocity scaling, the location of the energy peak associated with this mode can be predicted using similarity of the mean velocity in both the inner and core regions, in the form of a logarithmic profile in the overlap region and Reynolds similarity of the outer flow. The inner-scaled deviation of the centreline velocity from a log law is a constant, $C$. The relationship is
\begin{eqnarray}
U^+(y^+) = \frac{1}{\kappa}\ln y^+ + B,\\
U^+_{CL} = U^+(y^+=R^+)=\frac{1}{\kappa}\ln R^+ + B +C.
\end{eqnarray}
The velocity reaches a value of two-thirds of the centreline velocity at an inner-scaled wall-normal distance $y^+_{2/3}$ that can be predicted as follows:
\begin{equation}
U^+ = \frac{1}{\kappa}\ln y^+_{2/3} + B = \frac{2}{3} \left(\frac{1}{\kappa}\ln R^+ + B + C \right),
\end{equation}
or, rearranging,
\begin{equation}
y^+_{2/3} = a R^{+2/3}\label{eqn:y23}
\end{equation}
where $a=\exp[\frac{\kappa}{3}(2C-B)]$. For a pipe $\kappa = 0.421$, $B=5.60$ \citep{mckeonmean04} and $C\approx 2$ \citep{mckeon05}, so $a=0.8$.

\begin{figure}
\begin{center}
\epsfxsize=12cm
\epsfbox{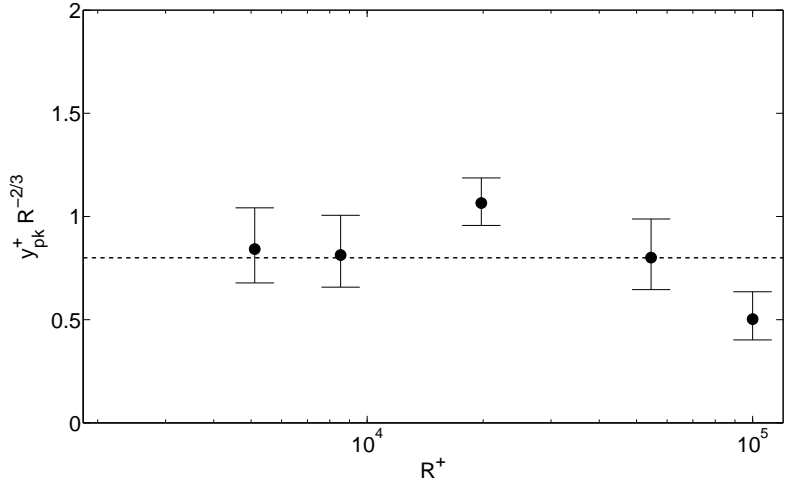}
\caption{Comparison of experimental pipe flow results on the $y^+$-location of the spectral peak in streamwise energy associated with the VLSM modes and the prediction of equation~\ref{eqn:y23}. $\bullet$: Superpipe data (Morrison \etal, 2004); dashed line: $y^+=0.8R^{+2/3}$. Note that the highest Reynolds number data point may have been affected by hot-wire spatial resolution (the non-dimensional wire length at this Reynolds number is $l^+ = 385$).}
\label{fig:VLSM_predict}
\end{center}
\end{figure}

The agreement is excellent with both the $(k,n)=(1,10)$ mode peak and the previously-published experimental variation of the VLSM streamwise energy peak shown in figure~\ref{fig:VLSM_predict}.  For the experimental data, the location of the VLSM peak at each Reynolds number was determined by considering the low wavenumber spectral peak in the hot-wire data set of \cite{Morrison04}, as reported in \cite{mckeonAIAA08}.  It should be noted that identifying the location of the peak energy relies on the local smoothness of the spectra.  Thus the error bars on the peak position shown in the figure are conservative and correspond to the next closest wall-normal locations at which spectra were obtained. While probe resolution effects may be a concern at the highest Reynolds number, the data at $R^+ = 19 \times 10^3$ clearly represent an outlier. For the remaining data the agreement is excellent. Thus we associate the first critical mode at this wavenumber pair, namely the critical mode with the lowest phase velocity, $(k,n,\omega) = (1,10,2/3 U_{CL})$, with the experimentally-observed VLSM, and infer that this mode gives a dominant energetic contribution to the turbulent fluctuations. Equation~\ref{eqn:y23} represents a non-observational attempt to predict the location of the peak in the streamwise energy associated with the VLSMs.

The scaling with the two-thirds power of Reynolds number is also reminiscent of critical layer arguments for neutrally-stable disturbances in linearly unstable laminar flow, thus we expand the critical layer framework to include forced, propagating modes in \emph{turbulent} pipe flow.

\subsection{A wall and critical layer framework}\label{sect:critlayer}

A consequence of the decomposition performed in Section~\ref{sect:approach} is that equation~\ref{eqn:linu} can be rewritten in terms of the vertical velocity and vorticity, yielding an operator equivalent to the more familiar Orr-Sommerfeld-Squire operator when the flow is linearised around the turbulent base flow, identified earlier as the $(k,n,\omega)=(0,0,0)$ mode. As such, we extend the tools of linear stability analysis to explore the scaling of the layers around the critical points, with reference to known results concerning the OSS equations.

To simplify the exposition, we consider the case for plane flow in Cartesian coordinates. Linearised plane flow has unstable (right-half plane) eigenvalues for sufficiently high Reynolds number and for the unforced case the ``neutral curve'' bounds the region of stability in the $Re-k$ plane. Pipe flow is always linearly stable, so has no neutral curve. However, comparable physical processes occur and are manifested as regions of high resolvent norm without actually reaching linear instability.
This said, we will work with the standard formulation of the Orr-Sommerfeld equation (O-S) for a plane flow to aid the discussion,
\begin{equation}
(U-\omega/k)(\curly{D}^2-K^2)\tilde{v}-U''\tilde{v}-\frac{1}{ik Re}(\curly{D}^2-K^2)^2\tilde{v}=0.
\label{eqn:OS}
\end{equation}
Here $U''$ is the second derivative of the base profile.

Scaling analysis of this equation reveals two wall-normal regions in which the action of viscosity is required. The singularity in the inviscid O-S (Rayleigh) formulation occurring if $\omega/k=U$ may be resolved by restoring either viscous effects or nonlinearities in a region local to the critical point.  We consider the case of viscous critical layers here, with justification to follow. Since the solutions of the Rayleigh equation do not obey the viscous boundary conditions, a second region in which the effects of viscosity are restored is required close to the wall. The scaling of the width of these regions can be determined by consideration of the appropriate terms in Equation~\ref{eqn:OS} (for plane flow see~\cite{Schmid01,Drazin81,Maslowe81}).

Close to the wall, the boundary layer approximation in a region around $y_1$ is given by
\[\frac{1}{ik Re}\curly{D}^4\tilde{v}=-\omega/k\curly{D}^2\tilde{v}=-U_x\curly{D}^2\tilde{v}\]
with solution
\begin{equation}
\tilde{v}(y) = C \exp\left(-(y-y_1)(i k Re U_x)^{1/2}e^{-i\pi/4}\right)
\label{eqn:visclayer}
\end{equation}
such that the viscous layer around $y_1$ has thickness of order $kRe^{1/2}$.

At the critical layer centred on $y_c$ the approximation to the O-S equation is
\[\frac{1}{ik Re}\curly{D}^4\tilde{v}=(U-\omega/k)\curly{D}^2\tilde{v} \approx U'_c(y-y_c)\curly{D}^2\tilde{v},\]
which reduces to an Airy equation for $V''$
\begin{equation}
\left(\dd{}{\xi}-\xi\right)\dd{}{\xi}V=0
\label{eqn:critlayer}
\end{equation}
under the substitutions $\epsilon = (ikReU'_c)^{-1/3}$, $\xi=(y-y_c)/\epsilon$ and $\tilde{v}(y)=V(\xi)$.  Therefore the critical layer thickness is of order $y/R \sim kRe^{-1/3}$ and the mode phase velocity, $U_x=\omega/k$, sets the wall-normal position of $y_c$.

Thus we expect to observe viscous and critical modes with $(kRe^{-1/2})$ and $(kRe^{-1/3})$ scaling, respectively, in the wall-normal direction.  Note that in inner scaling, this corresponds to $y^+ \sim (R^+/k)^{1/2}$ and $y^+ \sim (R^{+2}/k)^{1/3}$, respectively.

\begin{figure}
\centering
\begin{minipage}{0.45\linewidth}
    \begin{center}
\includegraphics[width = 7cm]{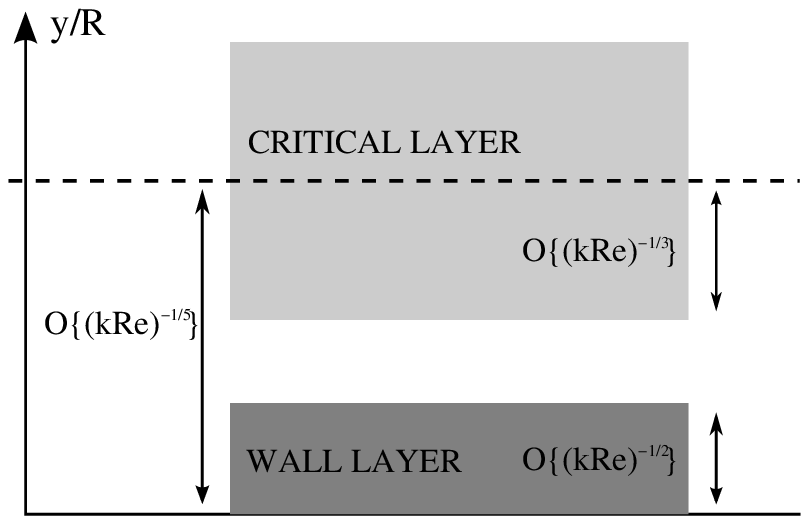}
\end{center}
\end{minipage}
\begin{minipage}{0.45\linewidth}
    \begin{center}
\includegraphics[width = 7cm]{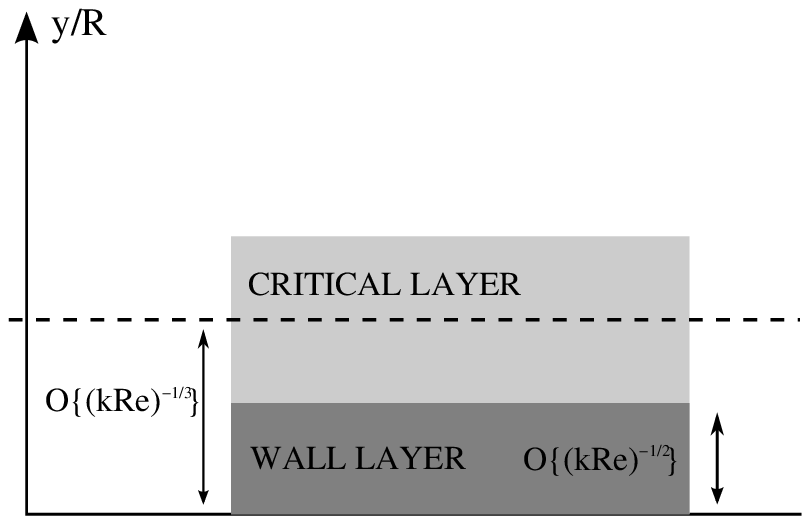}
\end{center}
\end{minipage}
\caption{Schematic of the scaling of upper (left) and lower (right) branch critical and wall layers in the Orr-Sommerfeld equation.}
\label{fig:criticallayers}
\end{figure}

The two options for the relative locations of the two layers are shown schematically in figure~\ref{fig:criticallayers}: if the layers are distinct (shown in the left panel) then the critical layer exists at a wall-normal distance ${\cal{O}}(kRe)^{-1/5}$ and if the critical layer reaches down to the wall, then it will be centred at a wall-normal location ${\cal{O}}(kRe)^{-1/3}$ (shown in the right panel).  In the O-S analysis, these two cases correspond to the upper and lower branches of the neutral stability curve at the Reynolds number under consideration.

A similar scaling analysis can be performed in turbulent pipe flow using the turbulent base profile and with an appropriate Reynolds number for turbulent flow, $Re=R^+$. There are no neutral curves because the eigenvalues remain in the left half plane for all Reynolds number and all wavenumber-frequency triplets. However similar effects to the true O-S system are manifested as high system response to forcing, giving a high resolvent norm with the same scaling. Hence we refer to upper and lower branch-type modes in what follows and consider the implications of wall and critical layers for the scaling of turbulent pipe flow and in particular the implied mechanism for the manifestation of viscous effects outside of the immediate near-wall region.

We conclude the this part of the discussion by observing that in laminar pipe flow the Reynolds-number independence of the form of the velocity profile means that there is a linear relationship between the point at which the velocity reaches two-thirds of the centreline value and the Reynolds number, $y^+_{2/3} \sim R^+$. It is known that the linear Navier-Stokes operator is particularly sensitive to perturbation at this point, where the wall and centre mode branches of the spectrum merge~\citep{Reddy93}.

\subsection{Non-normality versus criticality}

The foregoing analysis permits a comment on the importance of non-normality to the sensitivity to forcing. As indicated in Section~\ref{sect:approach}, the resolvent in Equation~\ref{eqn:resolvent} can become large in two distinct ways, via the non-normality arising from the gradient of the mean velocity, or when the diagonal terms tend to zero at a critical layer.
Regarding the non-normal mechanism, the local shear couples the streamwise and wall normal velocity components, leading to a ``lifting'' mechanism for streamwise vorticity.  On the other hand, the critical layer arguments arise from arguments that are normal in character, since this mechanism is present even when the mean shear is small. This provides a simple explanation of why critical modes are observed far from the wall in our model and become increasingly dominant at higher Reynolds number.  The decomposition performed in this work therefore permits a simple identification of the two effects.


\subsection{VLSMs as a consequence of critical layer scaling}

We predicted in Section~\ref{sect:VLSMpredict} that the wall-normal location of the peak energy of the mode associated with the very large scale motions scales with $R^{+2/3}$. The component-wise velocity distributions for the VLSM mode have been shown in figure~\ref{fig:modes2}. Many of the characteristics of this mode are in good agreement with those of the lower branch-type critical layer under the analysis described above. In turbulence terminology, the associated disturbance would be ``attached'' to the wall in the sense that it has finite amplitude in the near-wall region.  This property is consistent with the observations of the ``footprint'' of the VLSM modes reaching down to the wall (\cite{Hutchins07}, \cite{Guala09mod} and others) and influencing the instantaneous wall shear stress \citep{Marusictau07}.  This is a consequence of this mode retaining characteristics of the wall layer while having the lowest phase velocity which can be considered critical.  This phase velocity is a constant fraction of the centreline velocity, in agreement with the conclusions of del \'{A}lamo \etal~\nocite{delAlamoRetau04} that the centreline velocity is the appropriate velocity scale for the very long, ``global'' modes.

The mode shapes associated with the VLSMs in figure~\ref{fig:modes2} are in good agreement with the structure that can be inferred to give rise to the two-point correlations reported in the near-neutral surface layer by \cite{Guala09mod} and the conditional averages of \cite{Chung09} in channel flow. This suggests that the VLSM mode predicted here indeed becomes a dominant, viscous structure in the near-wall region as the Reynolds number increases.

The VLSM mode predicted here also has a concentration of fluctuation energy in the wall-parallel components, in agreement with the experimental evidence that the very long scales are visible in the $u$ and $w$ spectra (although the literature on the latter is relatively scarce due to the experimental difficulty of obtaining well spatially resolved information on the spanwise component), but not in the wall-normal component due to the effect of ``blocking'' due to the wall on eddy distribution. In addition, the spatial variations of the streamwise and wall-normal velocities implies a distribution of Reynolds shear stress that demonstrates the $\pi$ phase reversal associated with viscous critical layers in the vicinity of the peak in the streamwise energy. This justifies our earlier assumption that  viscosity is more important than nonlinearity in the vicinity of the critical layer. The phase relationships between the axial and wall-normal velocity components implies that these scales are ``active'' (in the sense of \cite{Townsend76}) in that they bear non-negligible Reynolds shear stress, as proposed by \cite{Guala06}, but in contrast to the spirit of Townsend's original ideas.

We note also that counter-rotating vortex structure implied by the VLSM mode shape (figure~\ref{fig:modes2}) is a well-known phenomenon associated with lower branch critical layer solutions, at least in the laminar case \citep{Viswanath09,Wang07}, and are observed at similar wavenumber pairs, $n \sim 10k$, using the current model for laminar flow \citep{Sharma09AIAA}.  The conditionally-averaged cross-stream streamlines and swirl distributions of \cite{HutchinsPTRSA07} and \cite{Chung09} confirm that this is the expected mode shape for the VLSMs in channel flows. This structure is also consistent with the work of del \'{Alamo} \& Jim\'{e}nez\nocite{delAlamo06}, who showed that the two types of disturbance that experience maximal transient growth in the initial value problem in turbulent channel flow, one inner scaling and one outer scaling mode, both resemble streamwise rolls and streaks.

\begin{figure}
\centering
\includegraphics[width = 12cm]{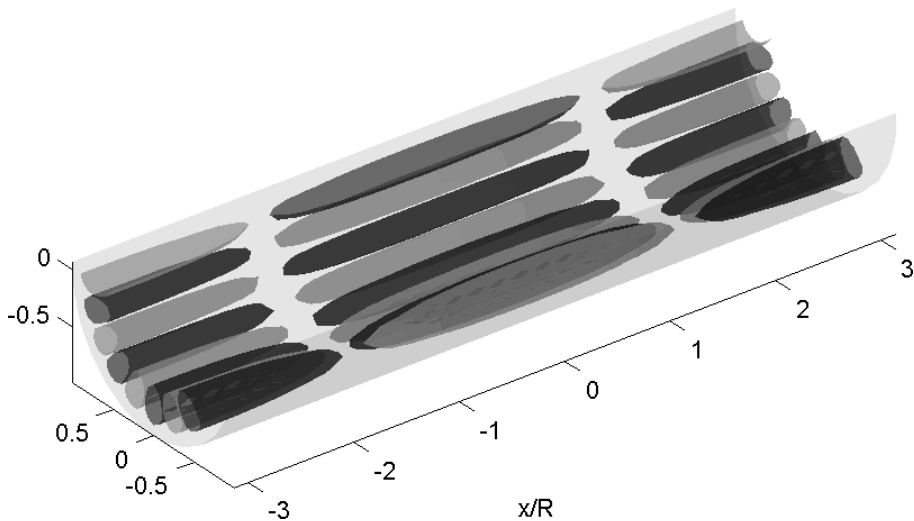}
\caption{Isocontours of streamwise velocity for the sum of left- and right-going VLSM propagating modes, $(k,n,\omega) = (1,\pm10, 1*2/3 U_{CL})$. The Reynolds number is $Re=75\times 10^3$.}
\label{fig:LR_sum}
\end{figure}

We conclude our comments on the relationship between VLSMs and critical layers by proposing a partial explanation of the apparent meandering, very long streamwise coherence of VLSMs that is indicated by experimental measurements of streamwise velocity in the spanwise plane for various flows.
To this end, figure~\ref{fig:LR_sum} shows isosurfaces of streamwise velocity for a sum of left- and right-going propagating VLSM-type modes. It seems plausible that the superposition of other modes on such a pair would reproduce something like the apparent observed meandering coherence of the order of 25$\delta$ \citep{Monty07}.

\subsection{The importance of wall layers}

Wall layer scaling has been implicit in the work of several researchers, including the experimental and theoretical discourses by \cite{Sreeni97}
and \cite{Sahay99} and the form of the mean velocity in the vicinity of this peak proposed by \cite{Sreeni06}. The proposed mean momentum balance structure of \cite{Klewicki07} describes the scaling in inner units with $R^{+1/2}$ of the Reynolds stress peak location. An analogy between the wall layer scaling phenomenon and linear stability concepts was drawn by \cite{Sreenivasan88} who explained the $R^{+1/2}$ dependence in terms of the roll-up of a vortex sheet used to model the vorticity in the boundary layer. \cite{Sirovich90} observed plane waves in channel flow travelling at the speed of corresponding to the mean velocity at the location of the Reynolds stress peak.

The foregoing discussion has given a rigorous derivation of the origin of the wall layer scaling. Such arguments can be extended to discuss local scaling of the turbulent spectrum. While it is understood that the turbulent fluctuations in the near wall region resist inner scaling, the cause of this discrepancy has been understood for some time to lie in the increase in energy at larger streamwise wavenumbers as the Reynolds number is increased, with the implication that the spectral shape at large $k$ (short wavelengths) has the potential to be self-similar.  In fact, the discrepancy is a consequence of wall layer scaling, which we now demonstrate.

Applying wall layer scaling for a mode with streamwise wavenumber $k$ gives
\begin{equation}
\frac{y_{pk}}{R} \sim \frac{1}{k R^{+1/2}}.
\end{equation}
For a constant $\lambda_x^+$, this can be rewritten in the following way, demonstrating that the peak energy location scaled in inner units will be independent of Reynolds number, provided that the mode phase velocity also remains constant:
\begin{equation}
y^+_{pk} \sim \frac{R^+}{\left(\frac{2\pi R^+}{\lambda_x^+} R^+\right)^{1/2}} = \left(\frac{\lambda_x^+}{2\pi}\right)^{1/2}.
\end{equation}
Based on the results shown earlier for $\lambda_x^+ \approx 1000$, where $y^+_{pk} \approx 13$, the constant of proportionality is ${\cal{O}}(1)$. So for the wavenumber range corresponding solely to wall layer arguments, the spectral shape should be self-similar in inner scaling. This was confirmed in figure~\ref{fig:lxpconst1}.
Predicting exact self-similarity would require further arguments relating to the Reynolds number dependence of the product of the forcing amplitude and the first singular value, which is beyond the scope of this paper.

\subsection{Amplitude modulation and the phase relationship between the large and small scales}

A region in the flow where an apparent amplitude modulation of the small scales occurs has been identified and explored by \cite{Blackwelder72,Bandy84,HutchinsPTRSA07,Mathis09} and \cite{Guala09mod}. This region can be identified with the concatenation of viscous layers at different wavenumber/frequency combinations and is required to meet the wall boundary conditions. Thus it must respond to the influence of finite amplitude critical layer disturbances. It may be said, then, that it truly represents an amplitude modulation of turbulence activity at the range of scales present in the near wall flow, the range of which is known to correspond to wavelengths smaller than the VLSMs. By the arguments made above, this viscous layer, in which the modulation is largest, should scale with $R^{+1/2}$.  This is borne out in the near-neutral atmospheric surface layer with $\delta^+ \sim {\cal{O}}(10^6)$ in the measurements of Guala \etal who show that the difference in energetic content of the small scales between positive and negative excursions of the fluctuation velocity in the VLSM bandpass-filtered signal is concentrated in the region $y^+ < {\cal{O}}(10^3)$.  In addition, the correlation coefficient between the large scale component and the filtered envelope of the small scale component of \cite{Mathis09} shows a distinct discontinuity in gradient that tracks a position $y^+ \sim \delta^{+1/2}$. We hypothesise that this corresponds to the outer edge of the inner viscous layer associated with the most energetic critical layer mode, which we identify as a VLSM.

\subsection{Implications for the scaling of the fluctuations in turbulent pipe flow}

Previous phenomenological observations and the predictions of the critical layer scaling can be concatenated into a skeletal description of turbulent pipe flow.  We showed above that wall layer scaling implies a similarity over the part of the turbulent spectrum governed by wall modes.  Additional arguments for the outer region of the flow (in which only critical modes contribute turbulent energy) seem to require the exact form of the mean velocity profile. However the similarity of the mode shapes for a constant velocity defect at high Reynolds number in figure~\ref{fig:kconst1} suggests that outer scaling of the spectrum can be recovered.  Therefore while outer fluctuations should be close to self-similar in outer units, the inner fluctuations will always include contributions from both wall and critical layer modes.  This, perhaps, is one cause of the apparent success of the ``mixed scaling'' proposed by \cite{DeGraaff}, which leads to collapse of $u^2/(u_\tau U_\infty)$ over a limited Reynolds number range. In light of the distinct footprints of the different classes of modes, the controversial question of the importance of ``top-down'' versus ``bottom-up'' effects may be more appropriately framed in terms of the relative importance of wall and critical layers, with the former tied to the latter through the wall boundary conditions, but with all fluctuations contributing to the shape of the mean profile.

The distributions of Reynolds shear stress associated with wall and critical modes have distinct shapes and, specifically, are local in $y^+$. That is, the Reynolds shear stress is also located in the perturbation layer, close to the peak energy. In addition, this localisation goes towards explaining some of the observations of \cite{Guala06} of the VLSMs, that these low wavenumbers (associated with critical modes in our formulation) are active in the sense that they contribute significantly to the Reynolds shear stress, and also that the sign of the wall-normal gradient in shear stress at this wavenumber, $\mathrm{d} (-uv(k))/\mathrm{d} y$, is opposite to that of the smaller scales closer to the wall. The reversal in sign of the mean Reynolds stress associated with a single mode lends support to the notion of a spatial transfer of turbulent energy away from the wall to the outer region, by simple arguments concerning the sign of the local turbulent energy production.  Of course, in the overlap region the integrated shear stress over all modes is such that local equilibrium concepts are at least approximately met.  There is a coupling between the sum of Reynolds shear stress contributions from all turbulent modes and the mean velocity profile that yields those modes.

For the VLSM mode, at least, a logarithmic variation in the mean velocity implies that this critical layer is always associated with a constant ratio of local to centreline velocity, $(U/U_{CL})^+$, with the consequence that knowledge of the location of the centre of the VLSM mode, which can be experimentally determined from the peak in the streamwise turbulent energy, yields Equation \ref{eqn:y23} relating the log law constants $\kappa,B$ and $C$. Alternatively, if the log law constants are known, then the peak offers an alternative way to determine the friction velocity, $u_\tau$.  While the footprint of the VLSM in the streamwise spectrum becomes more dominant as the Reynolds number increases, accurate identification of the location of the maximum becomes more difficult because the distribution around the peak also broadens in $y^+$, so these relationships may not prove to be useful in a predictive sense.

Our key assumption is that the modes indicated by the singular value decomposition will be observed in turbulent pipe flow if the wall-normal distribution of the forcing at that $(k,n,\omega)$ combination is non-zero, contains an exactly appropriate component and the singular value is large.  Clearly there are many other modes with the same streamwise wavenumbers or frequencies, as observed in the spatio-temporal spectra of, for example, \cite{Morrison69}, but different convective velocities which blur the harmonic decomposition.  However, \cite{Hutchins07} have used a synthetic signal to show that an isolated portion of a signal with clear streamwise and spanwise spatial content leads to a broad peak in the spatial power spectrum centred on the true signal wavenumbers.  We propose that the other modes combine to mask the signature of the VLSMs when a Fourier decomposition is performed,  such that the mode corresponding to the dominant VLSM critical mode is not clearly defined at $k=1$ in the power spectrum.

We do not address the spatial phase relationship between modes, however we suggest that the base of knowledge concerning inclined ramp-like structures and the wall-normal arrangement of statistically-representative hairpin vortices, together with recent work on the interaction between the large and small scales \citep{Mathis09, Chung09, Guala09mod} will inform future work in reassembling the modes predicted here to represent the essence of wall turbulence.

\subsection{Extension to other flows}

We have limited the arguments made above to turbulence in pipe flow.  However an equivalent formulation to that detailed in Section~\ref{sect:VLSM} can be made for other flows by considering the best values of the Karman and additive constants, and the wake deviation, for other canonical flows and retaining the assumption that the critical layer occurs when $U(y)=2/3 U_{CL}$ (i.e. that the critical layer scales with $y^+ \sim R^{+2/3}$).  For channel flow with $\kappa = 0.38$, $B=4.17$ and $C=0.175$ \citep{NagibChauhan08}, the estimation of the location of the peak associated with a lower branch-type critical VLSM is given by Equation~\ref{eqn:y23}, giving
\begin{equation}
y^+_{2/3} \approx 0.62 R^{+ 2/3}\label{eqn:model2}.
\end{equation}
The difference in the constant arises because of the form of the mean profile close to the centreline (the wake component), rather than the subtlety of the log law scaling, the potential (lack of) universality of which remains under debate~\citep{NagibChauhan08}. The selection of $k=1$ to represent the VLSM structure in channel flow is supported by the work of \cite{Monty09}, but that there is also significant energy observed at a streamwise wavelength of three channel half-heights, particularly in DNS studies.

\begin{figure}
\begin{center}
\epsfxsize=10cm
\epsfbox{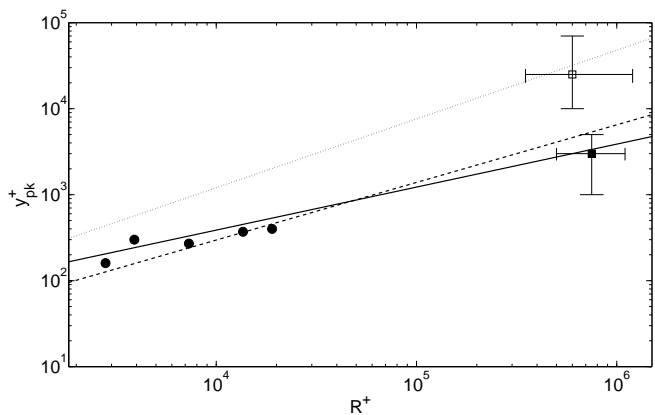}
\caption{Comparison of the $y^+$-location of VLSM-related phenomena in zero pressure gradient turbulent boundary layers over a range of Reynolds number. $\bullet$: location of zero amplitude modulation in laboratory boundary layers, estimates of the zero modulation location $\square$ (both Mathis \etal, 2009) and the VLSM energetic peak $\blacksquare$ (Guala \etal, in preparation) with various predicted forms.  --- $\sqrt{15}\delta^{+1/2}$ (Mathis \etal, also wall layer prediction), $- -$ $0.67\delta^{+2/3}$ (lower branch-type critical layer prediction), $\cdots$ $0.76\delta^{+4/5}$ (upper branch-type critical layer prediction).}
\label{fig:VLSM_predict_comp}
\end{center}
\end{figure}

A similar value for the constant, $a=0.67$ , is obtained by consideration of the log law constants in zero pressure gradient turbulent boundary layers.  The analysis for this flow requires accounting for, or formally proving negligible, the influence of spatial inhomogeneity in the streamwise direction, i.e. there is an implicit parallel flow approximation here. However there is utility in comparing the prediction of Equation~\ref{eqn:model2} with the experimental data.  The results are plotted in figure~\ref{fig:VLSM_predict_comp} which shows the experimentally-determined locations where the degree of amplitude modulation is zero in the data of \cite{Mathis09} (which they equate with the VLSM energy peak) and several predicted Reynolds number variations. Mathis \etal proposed $\delta^{+1/2}$ dependence of the peak location, after assuming that the VLSMs inhabit the exact middle of the overlap layer. We have shown earlier that it is also a consequence of wall layer scaling. The constants for the $\delta^{+4/5}$ scaling were determined by consideration of the point where the local velocity is four-fifths of the freestream value, in a similar analysis to Equation~\ref{eqn:y23}. While the $\delta^{+1/2}$ and $\delta^{+2/3}$ scalings could be said to describe the lower Reynolds number data equally well, only the $\delta^{+1/2}$ curve approaches one of the high Reynolds number data points obtained in the near-neutral atmospheric surface layer by Mathis \etal and Metzger \& McKeon.  The former study determined the higher point from a study with limited resolution close to the wall, while the latter study was performed with a maximum measurement height of the order of 10\% of the equivalent boundary layer thickness (and can be seen to reveal two spectral peaks). The critical layer framework provides an interesting interpretation of these two identified spectral energy maxima.

The data support a scaling with $\delta^{+1/2}$, particularly if the lower estimate of the VLSM peak energy in the surface layer is taken into account, suggesting that in the boundary layer there is a strong energetic contribution from the viscous wall mode (or, of course, the critical layer scaling changes due to the spatial inhomogeneity over the spatial VLSM period).  The $\delta^{+1/2}$ scaling is in agreement with the observation by Jimenez that the global modes travel with a convective velocity $U_s = 0.5 U_{CL}$ if the mean velocity is given by a log law governed by Equation~\ref{eqn:y23}.

However the error bars around the peak in the streamwise turbulent energy identified by Marusic \etal encompass the point $y^+ = 0.76 \delta^{+ 4/5}$, raising the intriguing possibility that we are observing the upper branch-type configuration of figure~\ref{fig:criticallayers}, with \emph{two} local maxima in the streamwise spectrum at $y^+$ locations proportional to $\delta^{+1/2}$ and $\delta^{+4/5}$. This may be a difficult hypothesis to prove, given the difficulty of obtaining fully-resolved spectral information for Reynolds numbers $20 \times 10^3 < \delta^+ < 7 \times 10^5$ or distinguishing between two spectral peaks that are close in physical and spectral space.  However we note that the streamwise velocity probability density functions reported by, amongst others, \cite{Adrian00} and \cite{Morris07} suggest two maxima that can be shown to correspond to approximately 0.5 and 0.8 times the free-stream velocity, which may confirm our hypothesis concerning the special significance of these velocities in the boundary layer.

The existence of upper and lower branch-type modes in the different canonical configurations of wall turbulence may contribute to the explanation for the emerging, subtle differences between the flows. In particular we note that the dominance of a different branch could affect the Reynolds stress profile with potential implications for the universality of even the mean velocity profile.

These results have assumed that the VLSM-type modes occur universally at $k=1$, while there is experimental evidence to suggest that the exact streamwise length varies from flow to flow, with the largest and smallest wavelengths observed in pipe and boundary layer flows, respectively.  However the framework presented above can easily be modified to account for this, should it prove to be required.

\section{Broader implications}

The critical layer arguments made here have several broader implications for canonical and more general flows. Next we identify some that are most pertinent to our current areas of study.

\subsection{Importance of the linear operator}
We have shown that consideration of the linear operator leads to a prediction for the dominant energetic mode shapes.  Note, however that a linearisation \emph{per se} is not strictly required.
This result is in agreement with the understanding that the linear operator alone is responsible for the extraction of energy from the mean flow. This suggests that the modes predicted by this analysis correspond more to the production and spectral content of energy at a particular wavenumber pair, with the full spectrum determining the rate of transfer between wavenumbers via nonlinear interactions.


\subsection{On Taylor's hypothesis}
The modal decomposition proposed in this work is consistent with the limited observations of the spatio-temporal spectra in the literature. The conversion between space and time is usually performed using Taylor's hypothesis of frozen turbulence, which is traditionally stated in terms of the slow evolution of the small-scale turbulence relative to the convective timescale and is believed to hold using the local mean velocity as the convection velocity for all but (possibly) the longest lengthscales \citep{Dennis08} and very close to the wall \citep{Morrison69}.  The foregoing analysis suggests an alternative interpretation: the critical layer analysis implies that the dominant modes (and, by implication, structures) travel with a phase velocity that corresponds to the local mean at the peak streamwise mode energy.  The reduction in amplitude away from the peak means that in the core part of the flow, the $(k-\omega)$ joint spectra should be narrow with a peak ridge lying on $\omega=U(y)k$.  Nearer the wall, the influence of the wall modes and the critical modes that extend to the wall but have energetic peaks further out will be to broaden the distribution of energy in $(k-\omega)$ space, remove the symmetry when integrating in $k$ or $\omega$ and ultimately lead to an inferred convective velocity that is different from the local mean velocity.

This suggests that issues related to the validity of Taylor's hypothesis would be better expressed in terms of the symmetry and width of the $k-\omega$ spectrum. The work of Morrison \etal\citep{Morrison69, Morrison71}, among others, illuminated the deviation of the inferred convective velocity from the local mean, the lack of symmetry near the wall in pipe flow and the spectral broadening that our model suggests. Recent Large Eddy Simulations in turbulent channel flow \citep{Chung09} reveal similar phenomena; this is an area of current experimental investigation.

Consideration of the variation of the spanwise wavenumber of energetic modes, the $n-\omega$ spectrum, suggests an additional source for the spatial resolution problems that hamper the majority of laboratory high Reynolds number wall turbulence experiments, namely the spanwise extent of the sensing element in viscous units.

\subsection{The minimal unit for turbulence}
The ``autonomous'' simulations of \cite{Jimenezlargescale04} demonstrated that a DNS of turbulent channel flow with a filter at a certain wall-normal height was capable of capturing the majority of the near-wall statistical features. This informed the ``top-down'' versus ``bottom-up'' debate and appeared to indicate that the external flow did not exert any control on the near-wall cycle.  In light of the critical layer analysis, we observe that the wall-normal cut-off in these simulations, $y^+ \sim 70$, is above the peak of the VLSM critical mode for the nominal Reynolds number of $Re_\tau \approx 550$, for which $y^+_{2/3} \sim 50$, so that even the lowest filter does not exclude the dominant, VLSM critical mode.  To this extent, the results are consistent with our hypothesis.  The implication for future simulations, then, is that the criterion on the minimum size of an autonomous simulation must be large enough to capture the azimuthal/spanwise and radial/wall-normal extent of the critical mode associated with the VLSM, or the furthest reaching mode that leads to considerable modulation in a viscous layer near the wall.

\subsection{Experimental development length}
A simple estimate for the development length required in canonical wall turbulence experiments can be made by assuming the importance, or at least the large singular values, associated with modes with a streamwise-spanwise aspect ratio of 10:1, as in the VLSMs explored here. The peculiar constraint on azimuthal symmetry in pipe flow requires that $n$ is integer, such that $k \ge 0.1$ for modes with the required aspect ratio. This is in good agreement with the observed low wavenumber end of the premultiplied spectra in the overlap region in pipe flow \citep{mckeoninertial07}, at a wavelength $\lambda_x \approx 60 R$.  Combining this figure with the estimates detailed in \cite{ZS} to account for the transition length at ($L/R < 14$) and the length for the shear layers to meet ($L/R \sim 60$) at $Re_D \sim 10^5$ gives a reasonable estimate for the minimum required development length, $L_D/R > 134R$. This figure is significantly exceeded in the Princeton Superpipe, where all measurements were taken for $L_D/R > 164R$. The estimate is in good agreement with the length $L_D \ge 160R$ for fully-developed statistics in pipe flow \citep{Monty01} or $L/R \approx 134$ in channel flow \citep{Dean76}, both at similar Reynolds numbers. This estimate also has implications for simulation, where box sizes can be set without consideration of the same entry effects experienced in experiments.

\subsection{Turbulence over rough walls}
The wall and critical layers indicated by our model suggest a physical explanation for several keystones of our understanding of rough wall flows.  Firstly, for flows that obey Townsend's hypothesis in the mean velocity, where the effect of the roughness is felt only in setting the friction velocity such that the self-similar form of the smooth wall is retained in the outer region, it is logical that the outer fluctuations would also obey similar scaling to the smooth wall case. This is because the outer region is dominated by modes whose shape is dictated by the local (i.e. outer) mean velocity profile.

Secondly, while it is known that the near-wall streamwise intensity peak is suppressed relative to smooth wall values, the modulation effect associated with the VLSMs is still strong over walls with $\nu/u_\tau \ll k< \delta$, for example in the near-neutral surface layer under conditions with $k_s^+ \le 50$ \citep{Guala09mod,Mathis09}.  In this flow, the VLSM critical layer is centred outside of the roughness sublayer of order $3k_s^+$. The difference in the wall boundary condition would suggest that the structure of the viscous layer required to meet the wall boundary conditions would be quite different from that observed in the smooth wall case, in that it would be significantly less coherent and a strong function of the roughness geometry. However an analysis similar to that given in section~\ref{sect:critlayer} could be performed on the new mean flow, still yielding a wall layer that must meet the boundary condition in response to the velocity field imposed by critical modes.  This is equivalent to stating that modification to the linear operator due to the roughness is confined to the region close to the wall, with the consequence that only the wall layer is significantly affected, for small enough roughness. Thus our critical layer model has the potential to make further progress in understanding the influence of surface roughness.

\subsection{Additional observations}
A full investigation of spectral scaling predictions using the critical layer model is reserved for future work. It also remains to cement the link between our model, statistical experimental data and the emerging picture of the structure of near wall turbulence. The picture has elements ranging from the VLSM structures discussed above through to observations of hairpin-like vortices, packets and uniform momentum zones \citep{Adrian07} and the near-wall cycle (previously thought to be autonomous) that is consistent with the physical arguments of \cite{Schoppa02}.  We also note that the critical layer arguments hold significance for approaches to control of wall turbulence, specifically suggesting the possible utility of using distributed (harmonic) forcing at the wall.

\section{Conclusions}

We have described here a resolvent norm model capable of describing the most amplified velocity modes in turbulent pipe flow, and an example of its application to illuminating scaling relationships in wall turbulence.

To derive the model, we treat the nonlinearity in the perturbation equation (involving the Reynolds stress) as an unknown exogenous forcing, yielding a linear relationship given by the resolvent, between the velocity field response and this nonlinearity. By only considering disturbances that are periodic in the wall-parallel directions and in time, we restrict our attention to propagating modes. This permits us to compare our results to the typical statistical decomposition of point and field experimental measurements and simulations for both temporal and spatial data.The linearity of the equation means that the different frequencies and wavenumbers are only coupled through the forcing.

Of necessity, there is a steady component that varies only in the wall-normal direction, which we identify as the turbulent mean profile. This yields for the perturbations a linear operator identical to that obtained by linearisation about the turbulent mean profile. However, we do not assume small perturbations, and such a linearisation would set the forcing to zero.

The mean profile is sufficient information to calculate the resolvent at any other wavenumber-frequency combination, and we use a mean profile determined from the Princeton superpipe experiments.

Once the resolvent is known, we find the forcing shape at a wavenumber-frequency combination that gives the largest velocity field response, under the assumption that it is likely to be a more  dominant structure in turbulent flows than other velocity field responses associated with smaller amplification.

The results of the model lend themselves to an interpretation, at each wavenumber-frequency combination, involving propagating modes with convection or phase velocity equal to the velocity somewhere in the interior of the flow.  Building on the analysis of the Orr-Sommerfeld equation in unstable flows, we identify two regions in which the action of viscosity is required for each wavenumber-frequency combination. They are in the vicinity of the critical layer, where the phase velocity is equal to the local mean velocity, and close to wall, where the boundary condition must be met.  We designated these the ``critical'' and ``wall'' layers. By analogy with the upper and lower branches of the neutral curve in unstable flow, there exist two relative wall-normal locations for these layers: an upper branch type configuration in which the critical layer resides at a distance $y^+ \sim R^{+4/5}$ and is distinct from the the wall layer which is centred on $y^+ \sim R^{+1/2}$, and a lower branch type solution in which the critical layer occurs at $y^+ \sim R^{+2/3}$, overlaps with the wall layer and has a footprint down to the wall.

It was proposed that key features of wall turbulence can be represented in the framework of a range of propagating modes with different wavenumber-frequency combinations such that they move with different convective velocities.
The dominant modes that emerge from the resolvent model provide a descriptive basis for the decomposition of the turbulent velocity field and capture some known features of wall turbulence.
In some sense, the conclusions of this research represent a return to the wave-like concepts of turbulence research in the mid twentieth century, but with the interpretation that coherent structure is a consequence of the assembly of modes and their relative motions.

While previous researchers have identified wall layer scaling for the Reynolds stress (Sreenivasan \etal, Klewicki \etal) and the importance of propagating waves (Sirovich \etal) in wall turbulence, our extended formulation suggests a picture of organised complexity in wall turbulence that appears to be consistent with a range of results from observations of full flows. In particular, the critical layer arguments and scaling that are presented here lead to inner scaling of the small scales near the wall and apparent outer scaling of the fluctuations in the core of the pipe, as well as a mechanism for the origin and scaling of the very large scale motions in pipe flow, as the critical layer solutions of the forced linear equation.  In addition, the interpretation of the wall layer as the solution which meets the wall boundary condition at each wavenumber-frequency combination implies that there must be an interaction between the critical and wall modes, an interaction that has been identified in recent literature as an amplitude modulation of the near-wall turbulence by the very large scales (Mathis \etal) or the phase relationship between the small and large scales (Chung \& McKeon). The broader implications of this scaling include a framework with which to interpret flow over rough walls, which can be considered as a modification to the boundary condition that the wall layer must meet. This is consistent with observations in the literature that while the near-wall flow is disturbed by the roughness, the VLSMs are still observed (e.g. Guala \etal, 2009).

A further observation concerns the appropriate scaling properties at the very large scales.  It is clear that the VLSM wavelength scales on the outer lengthscale, in our case the pipe radius. The critical layer scaling requires that the velocity scale should be the centreline velocity, $U_{CL}$, which is in agreement with arguments of Jim\'{e}nez \etal concerning the ``global'' modes in channel flow.  The model suggests that the dominant mode with streamwise and spanwise wavenumbers representative of the experimental observations of the VLSMs in pipe flow should travel with a phase velocity equal to two-thirds of the centreline velocity, such that the critical layer occurs where $U(y) = 2/3 U_{CL}$. This occurs within the overlap region in the mean velocity for all Reynolds numbers considered here, and as such represents a mechanism by which the effects of viscosity can extend well beyond the immediate vicinity of the wall. This is in agreement with several recent results including studies of the mean velocity profile and mean momentum equation.

The lack of information in this analysis on the relative phase between modes precludes immediate identification of structural features such as those we would expect from observations of the real flow, including hairpin-like vortices and the near-wall cycle, etc. However these may be construed, at least in part, to be a consequence of the dominant, energetic modes moving relative to each other with different convective velocities. This is a subject of ongoing investigation.

Extrapolation of the scaling arguments to other flows suggests a fundamental difference between internal and external flows (or at least pipes and boundary layers).  While the influence of spatial development in the latter case has not yet been taken into account, preliminary observations suggest that the upper branch type critical layer solution at VLSM-like a wavenumber pair may dominate, such that the VSLM energy peak scales with $y^+ \sim R^{+4/5}$ instead of the $R^{+2/3}$ scaling observed in pipe flow. Experimental confirmation of the critical layer scaling proposed here is ongoing.

Finally, the observation that the wall layer must meet the boundary condition suggests that the outlook is optimistic for wall-based control of friction drag.

\vspace{0.2in}

The support of NSF CAREER award grant number 0747672 (program managers William Schultz and H. Henning Winter), a PECASE award through the Air Force Office of Scientific Research (program manager John Schmisseur), grant number FA9550-09-1-0701(B.J.M.), an Imperial College Junior Research Fellowship and the EPSRC (A.S.) is gratefully acknowledged. We would also like to thank A. Meseguer and L. Trefethen for publishing their pipe code. 

\section*{Appendix A: Interpretation of resolvent analysis}\label{appendixa}

In addition to the type of transfer function analysis presented here, our resolvent analysis can be understood in terms of pseudospectra analysis \citep{Trefethen}.

The $\epsilon$-pseudospectrum of $\curly{L}$, $\Lambda_{\epsilon}$ is defined as

\begin{equation}
\Lambda_{\epsilon}(\curly{L}) = \left\{ \lambda \in \bspace{C}: (\lambda I-\curly{L})^{-1} \geq \epsilon^{-1} \right\}.
\end{equation}

That is, the $\epsilon$-pseudospectrum represents the spectrum of the operator
$\curly{L}$ under a perturbation with magnitude $\epsilon$.
This is then interpreted as bounds on the complex plane within which the spectrum of the perturbed operator lies.

The level curves of the pseudospectrum therefore correspond to the level curves of the resolvent and the spectrum of $\curly{L}$ represents the subset of the field of complex numbers where $\epsilon=0$ and the resolvent is unbounded.
The norm of the resolvent ($||(\lambda I-\curly{L})^{-1}||_2$) is equal to $\epsilon^{-1}$ at any particular point $\lambda$ in the complex plane.

\begin{figure}
\begin{center}
\epsfxsize=8cm
\epsfbox{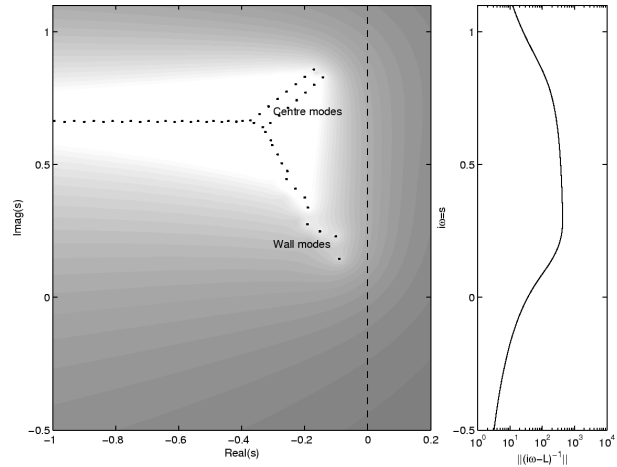}
\caption{Spectrum, pseudospectrum and resolvent norm for harmonic forcing, for a disturbance with $(k,n)=(1,10)$ in laminar pipe flow at $Re=5 \times 10^3$. Level curves at $(sI-\curly{L})^{-1} = 10^1, 10^2, 10^3,$ etc.}
\label{fig:pseudospecres}
\end{center}
\end{figure}

This is illustrated in the left pane of figure~\ref{fig:pseudospecres}, in which pseudospectrum of a representative mode in laminar pipe flow is compared with the familiar spectrum.
Where the pseudospectrum intrudes significantly on the right-half plane, significant perturbation growth with a nominally stable operator can result~\citep{Meseguer03}.

For pipe flow, the spectrum of $\curly{L}$ lies in the left-half plane ($\curly{L}$ is always stable). Since the maximum value of an analytic function on a region of the complex plane lies on that region's boundary (the maximum modulus principle), the maximal norm of the resolvent taken over the right half of the complex plane of a stable operator lies on the the imaginary axis. This corresponds to the system response to harmonic forcing and can be visualised by taking a slice of a pseudospectrum contour plot along the imaginary axis, as shown in the right-hand panel of figure \ref{fig:pseudospecres}.

Our resolvent analysis may be understood as an attempt to characterise system non-normality. Performing this characterisation by formulating the initial value problem is popular in fluid mechanics. The two approaches are related, but not trivially. It is a consequence of the Hille-Yosida theorem \citep{Curtain} that the resolvent obeys
\[\norm{(\lambda I-\curly{L})^{-1}}{2}  \leq \frac{M}{q-d}\] where $q$ is the real part of $\lambda$, and $M$ and $d$ are real numbers and
\[\norm{e^{\curly{L}t}}{2} = M e^{dt}.\]
$M$ characterises the transient behaviour, and $d$ the asymptotic response. It is an important open problem in control theory how to analytically determine $M$ for a given $d$ \citep{Blondel04}.  The properties of the resolvent in turbulent flow have been studied in the context of the initial value problem in wall shear flows \citep{Butler93,delAlamo06,Cossu09} and the response of turbulent channel flow to stochastic forcing \citep{Farrell98}.

\section*{Appendix B: Robustness of the resolvent norm to numerical and modelling error}\label{appendixb}

\begin{figure}
\begin{center}
\epsfxsize=8cm
\epsfbox{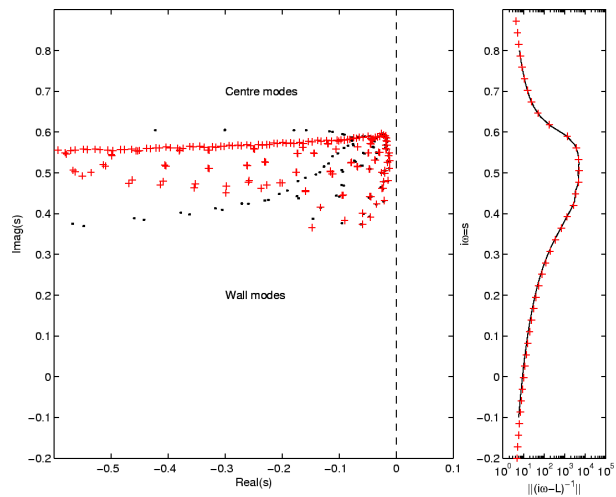}
\caption{Spectrum and resolvent norm of (turbulent) $\curly{L}$ for a disturbance with $(k,n)=(1,10)$ at $Re=7.5\times10^4$. In the left plane, the black dots show the spectrum calculated with a resolution of $N=60$ and the red crosses the same calculation for $N=200$. The right pane shows the resolvent norm across the imaginary axis. Although the spectrum is not well resolved overall in either case, the spectrum near the imaginary axis is well resolved quickly and the resolvent norm is accurate for much lower $N$ than is required for a full spectral analysis.}
\label{fig:turbspecres}
\end{center}
\end{figure}


Figure~\ref{fig:turbspecres} shows the spectrum of the operator $\curly{L}$ for the wavenumber pair $(k,n)=(1,10)$ in turbulent pipe flow for two different spatial resolutions, $N=60$ and $N=200$. We should expect that features prominent in real flows should be robust to operator perturbation (such as under-resolution) and as such have a high response to forcing. Inspection of figure~\ref{fig:turbspecres} demonstrates this to be the case.

In the left plane, the black dots show the spectrum calculated with a resolution of $N=60$ and the light crosses the same calculation for $N=200$. The right pane shows the resolvent norm across the imaginary axis. Although the whole spectrum is not well resolved overall at $N=60$, the spectrum near the imaginary axis is well resolved, suggesting that the dominant part of the spectrum is resolved with low $N$, and both the resolvent norm and the mode shape (not shown) are accurate for much lower $N$ than is required for a full spectral analysis.  However high resolution is required to observe near-wall, inner scaling modes at high Reynolds number. These two constraints may be incompatible and limit the extension of these results to higher Reynolds numbers without additional consideration of the numerical technique.


\end{document}